\documentclass[12pt,doublespace]{article}
\usepackage{hyperref,url,breakurl,float}

\usepackage[dvips]{graphics}
\usepackage{natbib,amsfonts,enumerate,amsmath,verbatim,graphicx}

\def\san#1#2#3{\renewcommand\arraystretch{0.4}\tabcolsep 0pt\begin{tabular}{c}
    $\scriptstyle{#1}$\\$\displaystyle{#2}$\\$\scriptstyle{#3}$\end{tabular}}

\begin{document}

% Title of paper
\title{Joint modelling of ChIP-seq data via a Markov random field model}
\author{Y. Bao$^1$, V. Vinciotti$^{1,\ast}$, E. Wit$^2$ and P. 't Hoen$^{3,4}$\\[4pt]
   {\em \small $^1$School of Information Systems, Computing and Mathematics, Brunel University, UK}\\
    {\em \small $^2$Institute of Mathematics and Computing Science, University of Groningen, The Netherlands}\\
    {\em \small$^3$Department of Human Genetics, Leiden University Medical Center, The Netherlands} \\
    {\em \small$^4$Netherlands Bioinformatics Centre, The Netherlands}
%\ast{veronica.vinciotti@brunel.ac.uk. To whom correspondence should be addressed.}
}
\date{}
\maketitle
\noindent {\it Abstract:}
Chromatin ImmunoPrecipitation-sequencing (ChIP-seq) experiments have now become routine in biology for the detection of protein binding sites. %protein-DNA interactions, DNA methylation, and histone modifications in vivo.
In this paper, we present a Markov random field model for the joint analysis of multiple ChIP-seq experiments. The proposed model naturally accounts for spatial dependencies in the data, %amongst neighboring pre-processed genomic windows,
by assuming first order Markov properties, and for the large proportion of zero counts, by using zero-inflated mixture distributions. In contrast to all other available implementations, the model allows for the joint modelling of multiple experiments, by incorporating key aspects of the experimental design. In particular, the model uses the information about replicates and about the different antibodies used in the experiments. %This leads to a better estimation of the variability in the data and, thus, to a more robust detection of enriched and differentially bound regions.
An extensive simulation study shows a lower false non-discovery rate for the proposed method, compared to existing methods, at the same false discovery rate. Finally, we present an analysis on real data for the detection of histone modifications of two chromatin modifiers from eight ChIP-seq experiments, including technical replicates with different IP efficiencies.

\section{Introduction}
\label{Introduction} ChIP-sequencing, also known as ChIP-seq, is a
well-known biological technique to detect protein-DNA interactions, DNA methylation, and histone modifications in vivo. ChIP-seq combines Chromatin ImmunoPrecipitation (ChIP)
with massively parallel DNA sequencing to identify all DNA binding
sites of a transcription factor or genomic regions with certain
histone modification marks.  %The ChIP process captures cross
%linked and sheared DNA-protein complexes using an antibody
%against a protein of interest. After decrosslinking of the protein-DNA
%complexes, the final DNA pool is enriched in DNA fragments
%bound by the protein of interest, but there are always random genomic
%DNA fragments piggybacking on the specific DNA fragments.
The final data produced by the experiment provide the number of DNA
fragments in the sample aligned to each location of the genome. From this,
the aim of the statistical analysis is to distinguish the truly
enriched regions along the genome from the background noise. Whereas conventional transcription factors, that bind directly to the DNA, show sharp peaks at the regions of enrichment, chromatin modifiers tend to have much broader regions of enrichment and do not follow a peak-like pattern. The latter cannot be captured by standard peak-calling algorithms and require more sophisticated statistical models. This is the focus of the present paper. %In this paper we present a statistical model that is particularly suited to the detection of extended regions of enrichment.

As regions of the genome are either bound by the protein in question or not, it is quite common to analyse such data using a mixture model. Here the observed counts are assumed to come either from a signal or from a background distribution. A number of methods have adopted this approach, with some differences in the distribution chosen for the mixture. \citet{spyrou09}, in their BayesPeak R package, adopt a Negative Binomial (NB) mixture model, with a NB distribution used both for signal and background. \citet{kuan11}, in their  MOSAiCS package, adopt a more flexible NB mixture model, where an offset is included in the signal distribution and this distribution itself is taken as a mixture of NBs. \citet{spyrou09} show evidence that a NB mixture model outperforms a
Poisson mixture model, such as the one used by Iseq (\citealp{mo12}). \citet{qin10}, in their HPeak implementation, suggest to use a zero-inflated Poisson
model for the background and a generalized Poisson distribution for the signal.  In this
paper we consider a more flexible framework by allowing a
zero-inflated Poisson or NB distribution for the background and
a Poisson or NB distribution for the signal component.

Another feature of ChIP-seq data on histone modifications is the spatial dependency of counts for neighbouring windows along the genome. This is mainly the result of a common pre-processing step, whereby the genome is divided  into bins of some ad-hoc length. It is quite common to consider fixed-width windows, although dynamic approaches have also been considered (\citealp{mo12}). The sum of counts within each window is subsequently considered for the analysis. As a result of this, true regions of the genome that are bound by the protein in question could be easily found to cross two or more pre-processed bins.
This issue has been addressed in the literature with the use of Hidden Markov Models (HMMs) (\citealp{spyrou09,mo12,qin10}).

With few exceptions, the methods developed so far are limited to the
analysis of single experiments, with the optional addition of a
control experiment. When technical replicates or biological
replicates are available, the standard procedure is to perform the
peak calling on each individual data set and then combine the
results by retaining the common regions. This process has inherent statistical problems, as pointed out by
\citet{bardet12} and \citet{bao13}. %the process of comparing two
%samples by merely overlapping the genomic coordinates of their
%respective individually called binding peaks has inherent
%statistical problems and leads to an underestimation of binding
%similarity.
Despite the recognition of the need for biological replicates for
ChIP-seq analyses (\citealp{tuteja09}) and despite the fact that
several normalization methods have been proposed for multiple
ChIP-seq experiments (\citealp{bardet12, shao12}), very few methods
have been developed that combine technical and biological
replicates at the modelling stage. This would allow to properly account for the variability in the data, leading to a more robust detection of enriched and
differentially bound regions. \citet{zeng13} extend MOSAiCS
(\citealp{kuan11}), by developing a mixture model for multiple ChIP-seq datasets:
individual models are used to analyse counts for each experiment and a final model
is considered to govern the relationship of enrichment among different samples.
\citet{bao13} build mixture models for multiple experiments, where replicates are modelled jointly by an assumption of a shared latent binding profile.  In \citet{bao13}, we show how such a joint modelling approach leads to a more accurate detection of enriched and differentially
bound regions and how it allows to account for the different IP
efficiencies of individual ChIP-seq experiments. The latter has
probably been the main reason why joint modelling approaches of
ChIP-seq data have rarely been considered so far.

In this paper, we combine all the aspects described above into a single model, by proposing a one-dimensional Markov random field model for the
analysis of multiple  ChIP-seq data. Our model can be viewed as a hidden
Markov model where the initial distribution is a stationary
distribution. As such, we follow the existing literature on the
use of hidden Markov models for ChIP-seq data in order to
account for the spatial dependencies in the data (\citealp{spyrou09,mo12}). In contrast to the existing
HMM-based methods, we propose a joint statistical model for ChIP-seq data, under general experimental designs. In particular, we discuss the case of technical and
biological replicates as well as the case of different antibodies and/or IP efficiencies associated to each experiment. The remainder of this paper is organized as follows. In Section
\ref{section:Methods}, we describe the Markov random
field model and its Bayesian Markov Chain Monte Carlo (MCMC)
implementation. In Section \ref{section:Simulation}, we perform a
simulation study to compare our method with two existing HMM-based
methods, BayesPeak and iSeq, as well as with the joint mixture
model of \citet{bao13}. A real data analysis on eight experiments
for the detection of histone modifications of two proteins, CBP
and p300, is given in Section \ref{section:Realdata}. In Section
\ref{section:Discussion}, we conclude with a brief discussion.
\vspace{-1cm}
\section{Methods}
\label{section:Methods}
\subsection{A joint latent mixture model and its limitations}\label{section:latent model}
The data generated by ChIP-sequencing experiments report the number of aligned DNA fragments in the sample for each position along the genome. Due to noise and the size of the genome, it is common to summarise the raw counts by dividing the genome into consecutive windows, or bins. Since the majority
of the genome is expected to be not enriched, we would generally expect that some bins are
enriched regions, with a lot of tags, %(possibly a 'peak' for TF binding),
and most other bins are not enriched, containing only few
tags. %(a small number of background bins may be 'read-riched' due to
%GC bias or other reasons).
This scenario is well suited to a mixture model framework.

Let $M$ be the total number of bins and $Y_{mcar}$ the counts in
the $m$th bin, $m=1,2,\cdots, M$, under condition $c$, antibody
$a$ and replicate $r$. In the ChIP-seq context, the condition $c$ stands
for a particular protein and/or a particular time
point, and $r= 1,\ldots,R_{ca}$ is the number of replicates for
antibody $a$ under condition $c$, with $a=1,\ldots, A$. It is well known how a different level of ChIP efficiency is associated to different antibodies and how different IP efficiencies have been observed also for technical replicates (\citealp{bao13}). The current setup allows to account for these effects in the joint statistical modelling of multiple ChIP-seq experiments, under a variety of common experimental designs. %This should be considered in the joint modelling of multiple ChIP-seq experiments.
The
counts $Y_{mcar}$ are either from a background population
(non-enriched region) or a from a signal population (enriched
region). Let $X_{mc}$ be the unobserved random variable specifying
if the $m$th bin is enriched ($X_{mc}=1$) or non-enriched
($X_{mc}=0$) under condition $c$. Clearly, this latent state  does
not depend on ChIP efficiencies. %Similarly to the model used in
%MOSAiCS for single experiments (\citealp{kuan11}),
As in \citet{bao13}, we define a joint
mixture model for $Y_{mcar}$ as follows:
\begin{eqnarray*}\label{eq:full_likelihoo_mixture}
Y_{mcar}& \sim & p_{c} f(y| \theta^{S}_{car})+(1-p_{c})f(y|\theta^{B}_{car}),
\end{eqnarray*}
where $p_{c}=P(X_{mc}=1)$ is the mixture portion of the signal
component and $f(y, \theta^{S}_{car})$ and $f(y,
\theta^{B}_{car})$ are the signal and background densities for
condition $c$, antibody $a$ and replicate $r$, respectively. Using
this model, the regions are  detected as enriched or not by
controlling the False Discovery Rate (FDR).

Since we divide the genome arbitrarily in fixed-size windows, it
is possible that a region in a certain cromatin state crosses two or more bins. As a consequence
of this, it is reasonable to assume that the enriched regions have
some Markov property. We have checked whether this is the case for
ChIP-seq data from a real study, after dividing the genome into 200bp fixed windows. We have detected the enriched regions using the
latent mixture model above at a $5\%$ FDR  and then calculated the
conditional frequencies for each region, given that the previous
region is enriched or non enriched. We denote these by $f_{1|1}$
and $f_{1|0}$,  respectively. % (e.g. $f_{1|1}$ is calculated by
%$\displaystyle\frac{\# (X_{m-1}=1, X_{m}=1)}{\# (X_{m-1}=1)}$).
Table \ref{Table:conditional_freq} shows that these two
frequencies are generally not equal. The conditional frequency of
a current bin being enriched given that the previous bin is
enriched is generally larger than the conditional frequency of a
bin being enriched given that the previous bin is not enriched.
As a further evidence of Markov properties, Figure
\ref{fig:markov_plot} plots the bin  counts $Y_m$ for a region of the genome, for one ChIP-seq experiment. On
the right, the plot shows the posterior probability of enrichment,
using a latent mixture model. This plot clearly shows regions of
consecutive enriched bins.  \\
TABLE \ref{Table:conditional_freq} and FIGURE \ref{fig:markov_plot} ABOUT HERE.

The issue of spatial dependencies in ChIP-seq data is often overcome in the literature by repeating the
experiments using some ad-hoc shift of regions, usually taken as
half of the original window size. In this paper, we propose a natural
extension of the mixture model in (\ref{eq:full_likelihoo_mixture}), which accounts for the
spatial Markov properties of the data. This is described in the next section.

\vspace{-0.7cm}
\subsection{A one-dimensional Markov random field model}
%In this section, we propose an extension of a mixture model to account for Markov properties. As explained before, we assume that
The number of reads in bin $m$, $Y_m$, is either drawn from a
signal or a background distribution. To simplify the notation, we will
omit the subscripts $c,a,r$ in this section.
The first issue is about the choice of the mixture distribution. Together with the general expectation that a large part of the genome is not bound by the protein in question, unmapped genome regions and insufficient sequencing depth, i.e. an
insufficient total number of reads, give rise to an excess of
zeros in the observed data. This forms part of the background
noise and gives us the motivation to use a zero-inflated
distribution to model the background. As the data is in forms of counts, it is natural to consider either a  Zero-Inflated Poisson (ZIP) or
a Zero-Inflated Negative Binomial (ZINB) distribution. That is, conditional on
the latent state $X_m$,
\begin{equation*}\label{eq:mixturePoisson}
Y_m|X_m=0 \sim  {\rm ZIP} (\pi, \lambda_0) \: {\rm or} \:{\rm ZINB}(\pi, \mu_0, \phi_0), \:\:\:\: Y_m|X_m=1 \sim {\rm Poisson} (\lambda_1) \: {\rm or} \:{\rm NB}(\mu_1, \phi_1),\nonumber
\end{equation*}
where the probability density function of the zero-inflated Poisson is given by:
\begin{eqnarray}\label{eq:ZIP}
ZIP(y|\pi, \lambda)=\left\{ \begin{array}{lr} (1-\pi)+\pi
\exp(-\lambda) & \text{ if } y=0, \\\\
\displaystyle\frac{\pi\exp(-\lambda)\lambda^y}{y!} & \text{ if } y>0.
\end{array} \right.
\end{eqnarray}
and the probability density function of the zero-inflated negative binomial is given by:
\begin{eqnarray}\label{eq:ZINB}
ZINB(y|\pi, \mu, \phi)=\left\{ \begin{array}{lr} (1-\pi)+\pi(\displaystyle\frac{\phi}{\mu+\phi})^{\phi} & \text{ if } y=0, \\\\
\displaystyle\frac{\Gamma(y+\phi)}{\Gamma(\phi)\Gamma(y+1)}(\frac{\mu}{\mu+\phi})^y(\frac{\phi}{\mu+\phi})^{\phi} & \text{ if } y>0.
\end{array} \right.
\end{eqnarray}
%where $\Gamma(y)=\int_0^{\infty} e^{-t}t^{y-1}dt$ is the Gamma function.

A zero-inflated model can be seen as a mixture model of a zero
mass distribution and a Poisson/NB distribution, so it can be
interpreted with the use of another latent variable which
represents the extra zeros in the background regions that a
standard Poisson/NB distribution cannot account for. If we denote
this inner latent variable by $Z_m$ and $P(Z_m=1|X_m=0)=\pi$, then conditional on $X_m$, we
have
\begin{eqnarray}\label{eq:mixturePoissonXandZ}
Y_m|X_m=0, Z_m=0 &\sim & 1 (y=0)\nonumber\\
Y_m|X_m=0, Z_m=1 &\sim & {\rm Poisson}(\lambda_0)\: {\rm or} \:{\rm NB}(\mu_0, \phi_0), \:\: Y_m|X_m=1 \sim  {\rm Poisson} (\lambda_1)\: {\rm or} \:{\rm NB}(\mu_1, \phi_1).\nonumber
\end{eqnarray}

The latent variable $X_m$, representing the binding profile, is assumed to satisfy one
dimensional Markov properties, that is,
\begin{eqnarray}\label{eq:MRF}
P(X_m=i|X_{-m})=P(X_m=i|X_{m-1}, X_{m+1}), i\in \{0, 1\}.
\end{eqnarray}
%i.e. the probability of a certain state at location $m$, given the
%state values at all the other locations is the same as the
%probability of that state at location $m$, given only the state values
%at the neighboring locations.
This give the classical factorization of the joint density
\begin{eqnarray}\label{eq:CMRF}
P(X_1, X_2, \ldots,
X_M)&=&\pi_0(X_1)\san{M-1}{\prod}{m=1}q_{X_m, X_{m+1}}\end{eqnarray}
in terms of the initial state distribution $\pi_0$ and transition
probabilities $q_{i,j}=P(X_{m+1}=j|X_{m}=i), i, j \in \{0, 1\}$. Unlike the model above, in this paper we use a  more natural representation of the
joint density of the latent states for a one-dimensional Markov random field model, namely:
\begin{eqnarray}\label{eq:SMRF}
P(X_1, X_2, \ldots, X_M)&=&\frac{\san{M-1}{\prod}{m=1}P(X_m,
X_{m+1})}{\san{M-1}{\prod}{m=2}P(X_m)}
\end{eqnarray}
where $P(X_m, X_{m+1})$ is the joint probability of $X_m$ and
$X_{m+1}$ and $P(X_m)$ is the marginal probability of $X_m$. In particular, we
have $P(X_m)=\displaystyle\sum_{x_{m+1}} P(X_m, X_{m+1}=x_{m+1})$.

When  the $X_m$ are binary variables, as in our case, we can further re-write the model
(\ref{eq:SMRF}) as {\small
\begin{eqnarray}\label{eq:SMRF_binary}
P(X_1, X_2, \ldots, X_M)%=\frac{\delta_{1,1}^{n_{1,1}}\delta_{1,0}^{n_{1,0}}\delta_{0,1}^{n_{0,1}}\delta_{0,0}^{n_{0,0}}}{\delta_{1}^{(n_{1,1}+n_{1,0}-I[X_1=1])}\delta_{0}^{(n_{0,1}+n_{0,0}-I[X_1=0])}}
= {\delta_1}^{I(X_1=1)}{\delta_{0}}^{I(X_1=0)}
{\Big(\frac{\delta_{1,1}}{\delta_{1}}\Big)}^{n_{1,1}}{\Big(\frac{\delta_{1,0}}{\delta_{1}}\Big)}^{n_{1,0}}
{\Big(\frac{\delta_{0,1}}{\delta_{0}}\Big)}^{n_{0,1}}{\Big(\frac{\delta_{0,0}}{\delta_{0}}\Big)}^{n_{0,0}},
\nonumber
\end{eqnarray}}
where
\begin{eqnarray}\label{eq:para in SMRF_binary}
n_{i,j}&=& \#\{X_m=i, X_{m+1}=j\}, \:\: \delta_{i,j}= P(X_m=i, X_{m+1}=j), i, j \in \{0, 1\}, m=1, \ldots, M-1 \nonumber\\
\delta_{1}&=&P(X_m=1)=\delta_{1,1}+\delta_{1,0}, \:\:\: \delta_{0}=P(X_m=0)=1-\delta_{1}, \:\:\delta_{0,1}=\delta_{1,0}=(1-\delta_{1,1}-\delta_{0,0})/2. \nonumber
\end{eqnarray}
%and $\delta_{0,1}=\delta_{1,0}=(1-\delta_{1,1}-\delta_{0,0})/2.$

One can show that this model satisfies (\ref{eq:MRF}), that is the model is a one-dimensional Markov random field model. And if we notice that the transition probabilities satisfy $q_{i,j}=\delta_{i,j}/\delta_{i}$, the model can be further written in terms of the transition probabilities $q_{i,j}$ as following
\begin{eqnarray}\label{eq:SMRF_transition} P(X_1, X_2,
\ldots, X_M)
&=&{\Big(\frac{q_{0,1}}{q_{0,1}+q_{1,0}}\Big)}^{I(X_1=1)}{\Big(\frac{q_{1,0}}{q_{0,1}+q_{1,0}}\Big)}^{I(X_1=0)}q_{1,1}^{n_{1,1}}q_{1,0}^{n_{1,0}}q_{0,1}^{n_{0,1}}q_{0,0}^{n_{0,0}}.
\end{eqnarray}

The most attractive property of this model is that the initial
state distribution under (\ref{eq:SMRF}) is the stationary
distribution. %, i.e. we always start from the stationary distribution.
This is different from BayesPeak \citep{spyrou09}.
Note also that the Ising model of \cite{mo12} has one parameter
less than the model presented here: this corresponds to assuming that $q_{1,1}+q_{0,1}=1$, which is an unnecessary assumption and it is not normally satisfied by the data.
\subsection{Parameter Estimation}\label{section:single}
To simplify the notation, we define $\tilde{q}_1=q_{1,1}$ and
$\tilde{q}_0=q_{0,1}$. These are the probabilities that the
current state of a bin is 1 (enriched) given that the state of the left
bin is 1 and 0, respectively. We denote with $R_{c}$ the number of replicates under condition $c$. For simplicity, we drop the subscript $c$ in what follows and we assume that the same antibody is used for all replicates under a particular condition, which is often the case in practice. A similar derivation applies to the more general case as well as to the more specific case of no replicates. Assuming a ZINB-NB mixture model (zero-inflated NB for the background and NB for the signal), we aim to estimate the parameters $\Theta=( \tilde{q}_0,\tilde{q}_1, \pi, \mu_0, \phi_0, \mu_1,\phi_1)$. The joint likelihood of this model given the latent states, $\textbf{X}$, the inner variables  $\textbf{Z}_1,
\ldots, \textbf{Z}_{R}$ and data $\textbf{Y}_1, \ldots, \textbf{Y}_{R}$, is given by
\begin{eqnarray}\label{eq:rep_full_likelihood_NB}
P(\textbf{X}, \textbf{Z},\textbf{Y}|\Theta)&=& P(\textbf{X}|\Theta)P(\textbf{Z}|\textbf{X}=0, \Theta)P(\textbf{Y}|\textbf{X}, \textbf{Z}, \Theta) \nonumber \\
 &\propto&{\Big(\frac{\tilde{q}_0}{\tilde{q}_0+1-\tilde{q}_1}\Big)}^{I(X_1=1)}{\Big(\frac{1-\tilde{q}_1}{\tilde{q}_0+1-\tilde{q}_1}\Big)}^{I(X_1=0)}
\tilde{q}_{1}^{n_{1,1}}(1-\tilde{q}_1)^{n_{1,0}}\tilde{q}_0^{n_{0,1}}(1-\tilde{q}_0)^{n_{0,0}}\nonumber\\
&\times& \san{{R}}{\prod}{r=1} \pi_r^{\Sigma_m I(X_m=0, Z_{mr}=1)}\times (1-\pi_r)^{\Sigma_m I(X_m=0, Z_{mr}=0)}\\
&\times&\san{R}{\prod}{r=1}\san{M}{\prod}{m=1}\Big[\frac{\Gamma(y_{mr}+\phi_{0r})}{\Gamma(\phi_{0r})\Gamma(y_{mr}+1)}
\Big(\frac{\mu_{0r}}{\mu_{0r}+\phi_{0r}}\Big)^{y_{mr}}\Big(\frac{\phi_{0r}}{\mu_{0r}+\phi_{0r}}\Big)^{\phi_{0r}}\Big]^{I[X_m=0, Z_{mr}=1]}\nonumber\\
&\times&\san{R}{\prod}{r=1}\san{M}{\prod}{m=1}\Big[\frac{\Gamma(y_{mr}+\phi_{1r})}{\Gamma(\phi_{1r})\Gamma(y_{mr}+1)}\Big(\frac{\mu_{1r}}{\mu_{1r}+\phi_{1r}}\Big)^{y_{mr}}
\Big(\frac{\phi_{1r}}{\mu_{1r}+\phi_{1r}}\Big)^{\phi_{1r}}\Big]^{I[X_m=1]}.\nonumber
\end{eqnarray}
Here we assume that technical and biological replicates share the same
binding profiles, i.e. the latent states $X$ are common between replicates. This results in the joint probabilities $P(X_m, X_{m+1})$ in equation
(\ref{eq:SMRF}) being equal for all replicates, and consequently,
the transition probabilities $\tilde{q}_0$ and
$\tilde{q}_1$ in equation (\ref{eq:SMRF_transition}) are also equal across
replicates. A similar derivation applies for the ZIP-Poisson mixture model (zero-inflated Poisson for the background and Poisson for the signal) for the estimation of the parameters $\Theta=(\tilde{q}_0,\tilde{q}_1,  \pi, \lambda_0, \lambda_1)$.

We use a Bayesian methodology, in a Metropolis-within-Gibbs procedure,
to estimate the model parameters and states. In particular, we use a Direct Gibbs (DG) method to draw the latent state $X$.
DG treats each state as a separate parameter term and draws each
$X_m$, for $m=1, \ldots, M$, from its full conditional distribution
$$P(X_m=i|X_{-m}, \textbf{Y}_1, \ldots, \textbf{Y}_{R}, \Theta) \propto q_{X_{m-1}, i}q_{i, X_{m+1}}\san{R}{\prod}{r=1} P_i(Y_{mr}|\Theta)$$
where $X_{-m}=\{X_1, \ldots, X_{m-1}, X_{m+1}, \ldots, X_M\}$,
$P_i(Y_{mr}|\Theta)=P(Y_{mr}|X_m=i, \Theta)$ and the normalising constant is the sum over all possible values of $i$. Given $X_m=0$, the inner latent variable $Z_{mr}$ is drawn from its full conditional distribution %for all $m$ that
%$X_m=0$,
$$P(Z_{mr}=i|X_{m}=0, Y_{mr}=y_{mr}, \Theta) \propto P(y_{mr}|X_m=0, Z_{mr}=i, \Theta)P(Z_{mr}=i|X_m=0).$$
%where the normalising constant is the sum over all possible values of $i$.
We choose Gamma and Beta conjugate priors for the
parameters and draw samples from their posterior distributions. More
details are given in the supplementary material.

\subsection{Assuming the same number of binding sites across conditions}\label{section:same p}
The method above can be used in the presence of technical and biological replicates. Whereas technical and biological replicates share the same binding profile \textbf{X}, different proteins will generally have a different binding profile. Under certain conditions, e.g. when comparing the binding profiles across two conditions or between highly similar transcription factors, we can assume that the total number of binding sites is the same.
In \citet{bao13}, we show how this assumption  can  be included in a mixture modelling framework. % and how it helps to account for the different IP efficiencies of individual experiments.
In this paper, we show how the same assumption can be included also in the proposed Markov random field mixture model.

In particular, if $\textbf{X}_1$ and $\textbf{X}_2$ are the binding profiles of conditions 1 and 2, respectively (e.g. protein 1 and protein 2), we can include in the joint model the a priori biological
knowledge that the two conditions have the same number
of binding sites, i.e. $P(X_{m1}=1)=P(X_{m2}=1)$ for any region $m$. If the transition
probabilities $q$ are the same for the two conditions, then from our
stationary random field model (\ref{eq:SMRF_transition}), the
stationary distributions of the two experiments are also the same. %that is $P(X_{m1}=1)=P(X_{m2}=1)$.
However, assuming equal transition probabilities is quite a strong assumption and it is difficult to
know this beforehand, unless we are in the presence of technical
replicates. If we note that the stationary distribution
$P(X=1)=\displaystyle\frac{\tilde{q}_0}{\tilde{q}_0+1-\tilde{q}_{1}}=\frac{1}{1+(1-\tilde{q}_1)/\tilde{q}_0}$, we can
see that if
$\displaystyle\frac{1-\tilde{q}_{11}}{\tilde{q}_{01}}=\displaystyle\frac{1-\tilde{q}_{12}}{\tilde{q}_{02}}$
then $P(X_1=1)=P(X_2=1)$. Here $\tilde{q}_{01}$, $\tilde{q}_{11}$
and $\tilde{q}_{02}$, $\tilde{q}_{12}$ denote the transition
probabilities corresponding to the binding profiles $\textbf{X}_1$ and $\textbf{X}_2$ of the two
conditions, respectively. This shows that a weaker constraint in the transition probabilities is necessary to achieve equal probabilities of enrichment.

In general, let $s=\displaystyle\frac{1-\tilde{q}_{1c}}{\tilde{q}_{0c}}$ for protein $c$ and assume that $s$ is common for all proteins $c$, with $c=1, \ldots, C$, that is the different proteins have roughly the same number of binding sites. If $R_c$ is the number of replicates for protein $c$, then the joint likelihood given the latent states $\textbf{X}_1,
\ldots, \textbf{X}_{C}$, $\textbf{Z}_{11}, \ldots, \textbf{Z}_{1R_{1}},\ldots, \textbf{Z}_{C1}, \ldots, \textbf{Z}_{CR_{C}} $, and the data $\textbf{Y}_{11}, \ldots, \textbf{Y}_{1R_{1}}, \ldots, \textbf{Y}_{C1}, \ldots, \textbf{Y}_{CR_{C}}$,
is given by:
\begin{eqnarray}\label{eq:samep_full_likelihood}
&&\san{C}{\prod}{c=1}{\Big(\frac{1}{1+s}\Big)}^{I(X_{1c}=1)}{\Big(1-\frac{1}{1+s}\Big)}^{I(X_{1c}=0)}(1-s\tilde{q}_{0c})^{n^c_{1,1}}(s\tilde{q}_{0c})^{n^c_{1,0}}\tilde{q}_{0c}^{n^c_{0,1}}(1-\tilde{q}_{0c})^{n^c_{0,0}}\nonumber\\
&\times&\san{R_{c}}{\prod}{r=1}\pi_{cr}^{\Sigma_m I(X_{mc}=0, Z_{mcr}=1)}\times (1-\pi_{cr})^{\Sigma_m I(X_{mc}=0, Z_{mcr}=0)}\nonumber\\
&\times&\san{R_{c}}{\prod}{r=1}\san{M}{\prod}{m=1}\Big[\frac{\Gamma(y_{mcr}+\phi_{0cr})}{\Gamma(\phi_{0cr})\Gamma(y_{mcr}+1)}
\Big(\frac{\mu_{0cr}}{\mu_{0cr}+\phi_{0cr}}\Big)^{y_{mcr}}\Big(\frac{\phi_{0cr}}{\mu_{0cr}+\phi_{0cr}}\Big)^{\phi_{0cr}}\Big]^{I[X_{mc}=0, Z_{mcr}=1]}\nonumber\\
&\times&\san{R_{c}}{\prod}{r=1}\san{M}{\prod}{m=1}\Big[\frac{\Gamma(y_{mcr}+\phi_{1cr})}{\Gamma(\phi_{1cr})\Gamma(y_{mcr}+1)}\Big(\frac{\mu_{1cr}}{\mu_{1cr}+\phi_{1cr}}\Big)^{y_{mcr}}
\Big(\frac{\phi_{1cr}}{\mu_{1cr}+\phi_{1cr}}\Big)^{\phi_{1cr}}\Big]^{I[X_{mc}=1]}.\nonumber
\end{eqnarray}
This is used in a Metropolis-within-Gibbs procedure similar to the one described in the previous section and with more details provided in the supplementary material.
%As before, we sample the latent variable $X_{mc}$ from its full conditional distribution for $m=1, \ldots, M$, $c=1, \ldots, C$,
%$P(X_{mc}=i|X_{-mc}, \textbf{Y}_{c1},\ldots, \textbf{Y}_{cR_c},\Theta) \propto q_{X_{m-1,c}, i}q_{i, X_{m+1,c}}\san{R_{c}}{\prod}{r=1}P_i(Y_{mcr}|\Theta).$
%Given $X_{mc}=0$, the inner latent variable $Z_{mcr}$ is drawn using
%$P(Z_{mcr}=i|X_{mc}=0, Y_{mcr}=y_{mcr}, \Theta) \propto P(y_{mcr}|X_{mc}=0, Z_{mcr}=i, \Theta)P(Z_{mcr}=i|X_{mc}=0)$
%where the normalising constant is the sum over all possible values of $i$.
%Details about the posterior distributions of the parameters are given in the supplementary material.

\subsection{Identification of enriched regions and differentially bound regions}
In this section, we show how the statistical model described above is used to detect the regions in the genome that are bound by a protein of interest.
Let ${\bf X^{(1)}_{c}, \ldots, X^{(N)}_{c}}$ be $N$ Gibbs draws of
${\bf X_{c}}$ with ${\bf X^{(k)}_{c}}=(X^{(k)}_{1c}, \ldots,
X^{(k)}_{Mc})$. ${\bf X_c}$ is defined in section
\ref{section:latent model} and denotes the latent binding profile under
condition $c$. Under the proposed random field model, a natural
estimate of the posterior probability that the $m$th bin is
enriched is given by $\hat P(X_{mc}=1|{\bf Y})=\san{N}{\Sigma}{k=1} I(X^{(k)}_{mc}=1)$ (\citealp{scott02}).
To decide
whether a bin is enriched or not, we set a threshold on these probabilities based on the false discovery rate.
%Different criteria can be used
%to set this cut-off. In \citep{spyrou09}, an 0.5 cut-off
%is used, whereby each region is assigned to the state with the
%highest posterior probability. In this paper, as in
%\citet{broet06}, we use a cut-off corresponding to a specific
%value of the expected posterior false discovery rate.
If $D$ is
the number of enriched regions corresponding to a particular
cut-off on the posterior probabilities, then the expected false
discovery rate for this cut-off is given by $\overline{FDR} = \dfrac{\displaystyle\sum_{\rm m \: enriched} \hat P(X_{mc}=0|{\bf Y})}{D}.$

When data are available for more than one protein, the interest is also on finding the regions that are bound only by one of the proteins.
Following the idea of \cite{bao13}, we define a probability of
differential binding by
\begin{equation*}\label{eq:prob of differential binding}
P(X_{m1}\neq X_{m2})=P(X_{m1}=0|{\bf Y}_{1})P(X_{m2}=1|{\bf Y}_{2})+P(X_{m1}=1|{\bf Y}_{1})P(X_{m2}=0|{\bf Y}_{2})
\end{equation*}
where $P(X_{mc}=0|{\bf Y}_{c})=P(X_{mc}=0|{\bf Y}_{c1}, \ldots, {\bf Y}_{cR_{c}})$ is the posterior probability that the $m$th bin is enriched for protein $c$, estimated by the formula described above and from all the data on protein $c$. In this way, all technical replicates under the same condition are considered in the estimation of the posterior probabilities, returning a more robust set of differentially bound regions.

\section{Simulation study}
\label{section:Simulation}

In this section, we perform an extensive simulation study where we compare our method with three competitive methods: iSeq
(\citealt{mo12}), BayesPeak (\citealt{spyrou09}) and the mixture model approach of \citet{bao13}. For a number of different scenarios, we generate the data on
$M=10000$ regions and we repeat the simulation for 100 times. We use the Markov Random Field model (MRF) proposed in this paper, iSeq and
BayesPeak to estimate the parameters and to identify the enriched
regions by controlling the False Discovery Rate (FDR) at 0.05. We then
compute the False Non-discovery Rate (FNDR), that is the fraction of all the non-discovered regions that were actually enriched.  Finally, we use a t-test to test if the FNDR of
our method is significantly less than the FNDR of the two other methods and report the p-values.

In the first simulation study, we compare our method with other HMM-based methods, namely  iSeq (\citealt{mo12}) and BayesPeak (\citealt{spyrou09}). We consider four different scenarios. In the first scenario, we simulate data from a mixture model with a ZINB background distribution and a NB signal distribution. We set the parameters of these distributions using the values estimated by a MRF model on two of our real ChIP-seq datasets. We choose the experiments on the basis of their ChIP efficiency. In particular, we consider the case of a not very efficient experiment (CBPT0) and the one of a more efficient experiment (p300T302). In terms of the mixture distribution, the more efficient experiment corresponds to a background and signal distributions that are better separated. Since neither iSeq nor BayesPeak can deal with multiple experiments, we perform these comparisons on single experiments. The results are given in Table \ref{Table:simulation1} (scenario 1). BayesPeak is in general inferior to both iSeq and MRF. Between iSeq and MRF, there is no significant difference for the less efficient experiment, whereas MRF is superior to iSeq for the more efficient experiment. In general, we find that the use of zero-inflated distributions is particularly suited to the case of efficient experiments, where there is combination of a large number of zeros and a relatively large number of high counts. A mixture of Poisson distributions, which is implemented in iSeq, cannot capture this situation very well.
We further extend this simulation to scenarios where some assumptions are shared between MRF and iSeq. Firstly, we generate data from a mixture of Poisson distributions and  $\tilde{q}_1+\tilde{q}_0=1$. These are the two main assumptions imposed by the Ising model implemented in iSeq. As before, we simulate parameters for a more efficient and a less efficient experiment.  The results are given in Table \ref{Table:simulation1} (scenario 2). In this case, as expected, there is no difference between iSeq and MRF, whereas BayesPeak is still inferior to both. Secondly, we consider the case of a Poisson mixture, as in iSeq, but we relax the assumption of  $\tilde{q}_1+\tilde{q}_0=1$ (Table \ref{Table:simulation1}, scenario 3). In both cases, the MRF method is superior to iSeq, although the difference is not so large.
Finally, in the fourth scenario, Table \ref{Table:simulation1} (scenario 4), we generated data which satisfies the constraint $\tilde{q}_1+\tilde{q}_0=1$, but which does not follow a Poisson mixture distribution. In particular, we use a ZINB-NB mixture distribution. This is the case where the MRF method performs much better than either of the two other methods. In general, the results in Table \ref{Table:simulation1} show how iSeq and MRF perform equally well when the data is generated from a Poisson mixture distribution and $\tilde{q}_1+\tilde{q}_0=1$, but MRF is superior to both iSeq and BayesPeak when either of these two conditions is not satisfied.
TABLE \ref{Table:simulation1} ABOUT HERE.

In a second simulation study, we compare the MRF model with our previously developed mixture model for multiple experiments (\citealp{bao13}). For a fairer comparison, we now extend this model to include zero-inflated distributions for the background.
In particular, we test the performance of the two methods on data generated with and without Markov properties. Once again we consider the case of a very efficient experiment and the case of a less efficient experiment, and we generate two replicates in each case. The results are reported in Table \ref{Table:simulation2}. In the first scenario (scenario 5), the data are generated from a MRF model, using the parameter values estimated from two real datasets. As expected, the MRF model performs better than the mixture model in this case, as it accounts for the Markov dependencies. In the second scenario (scenario 6), we generated data without Markov properties, that is we generated the latent state $X$ simply by using a Bernoulli distribution. In this case, the MRF and mixture model give the same results.
TABLE \ref{Table:simulation2} ABOUT HERE.\\
From both simulation studies, one can conclude that the proposed MRF model performs as well as the other methods under similar conditions, but it outperforms the other models under more general mixture distributions and modelling assumptions.
\vspace{-1cm}
\section{Real data analysis}
\label{section:Realdata}
In this section, we use the new model on real ChIP-seq data on two proteins, p300 and CBP (CREB-binding protein). These are transcriptional activators whose regulatory mechanisms are not fully understood, but are thought to be quite crucial for a number of biological functions.
We analyze ChIP-seq data from six experiments, three for CBP and
three for p300 \citep{ramos10}. For each protein, one experiment is conducted at time point 0 (T0) and two technical replicates are performed after 30 minutes (T301 and T302). We also use CBP and p300 ChIP-seq
data from an earlier study \citep{wang09}, where CBP and p300 binding was evaluated in resting cells. The data are further described in (\citealp{bao13}), where we also discuss the effect of the different IP efficiencies on the resulting data. This is the case also for the technical replicates, with one replicate having a higher IP efficiency than the other. %All sequence reads were aligned to the human genome (build hg18) using BWA version 0.5.9 with default settings.
We divide the whole genome into 200 base pair windows and summarise the raw counts for each window by the
number of tags whose first position is in the window. The window length was chosen as the fragment size used in the ChIP-seq experiment. %As such, this is thought to be the smallest region that could be possibly found enriched.
Furthermore, we exclude from the analysis genomic regions that have been found to exhibit anomalous or unstructured read counts from the analysis (\burl{http://hgdownload.cse.ucsc.edu/goldenPath/hg18/encodeDCC/wgEncodeMapability/wgEncodeDukeRegionsExcluded.bed6.gz} \citep{hoffman12}).

First of all, we have compared the fit of a NB mixture model, where a NB distribution is chosen both for the background and signal, against a ZINB-NB mixture, where a zero-inflated NB is considered for the background and a NB distribution for the signal. In Figure \ref{fig:BIC}, we give the BIC values for the eight experiments, where we do not consider Markov properties. In general, we find that the BIC values are lower for the ZINB-NB mixture than for the NB mixture, suggesting a better fit for the zero-inflated mixture. In the following, we will therefore use zero-inflated distributions, differently to iSeq (\citealp{mo12}) and BayesPeak (\citealp{spyrou09}), which use Poisson and NB mixtures, respectively.
FIGURE \ref{fig:BIC} ABOUT HERE.

Within the eight data sets that we analyzed, CBPT0, p300T0, WangCBP and
Wangp300, are single experiments, i.e. with no replicates. We therefore
compare the proposed MRF model with iSeq and BayesPeak on these four
experiments. Table \ref{Table:realdata1}
gives the number of enriched regions identified by the three methods, respectively. For simplicity we provide the results only for chromosome 21. At the same 5\% FDR, MRF can detect more regions than any of the other two methods.
The overlap for all three methods is shown in the last row of the table, whereas pairwise comparisons are shown in the Venn diagrams  in Figure \ref{fig:venndiagram} for two representative experiments (CBPT0 and p300T0). In general, MRF tends to agree more with iSeq than with BayesPeak in the sense that MRF identifies most of regions that also identified by iSeq. This is consistent with what we observed in the simulation study, where we also showed a lower false non-discovery rate for MRF. Furthermore, Figure \ref{fig:venndiagram} shows how the overlap between MRF and iSeq and the overlap between MRF and BayesPeak are both much bigger than the overlap between iSeq and BayesPeak.
TABLE \ref{Table:realdata1} and FIGURE \ref{fig:venndiagram} ABOUT HERE.

CBP and p300 both have largely overlapping roles in transcriptional activation. We use {\tt ChromHMM} \citep{ernst10} to explore whether the regions identified by MRF are likely functional
in transcription activation and whether different chromatin features are enriched in the regions identified by the different methods. Figure \ref{fig:ChromHMM_1} shows the results of {\tt ChromHMM} using a 4-state hidden Markov model on the enrichment profile given by the three methods, each at a 5\% FDR, for two representative experiments (CBPT0 and p300T0). For each method, the data from the two proteins is jointly modelled by {\tt ChromHMM}. The left plots give the emission probabilities for the different analyses, that is the probability of the observed enrichment given each of the four possible states. These plots show how, for all analyses, three of the four states explain most of the enrichment pattern in the identified lists. The right plots give the relative fold enrichment for several annotations. These plots show how these three states are represented by a similar enrichment of features for the three methods, mainly CpGisland and RefSeq Transcription Start Sites (RefSeqTSS),  suggesting that the additional regions detected by MRF are likely to be genuine binding events.
FIGURE \ref{fig:ChromHMM_1} ABOUT HERE.

For replicated experiments (CBPT301, CBPT302, p300T301, p300T302), we compare the proposed MRF model for multiple experiments, with our previously developed mixture model (\citealp{bao13}). In both cases, we detect the enriched regions and the differentially bound regions, that is the regions bound only by one of the two proteins, at a 5\% FDR. Table \ref{Table:realdata2} reports the results for chromosome 21. The results show that by including the assumption of Markov properties, the number of enriched regions detected is larger than when the Markov property is not considered. This is to be expected since regions with a relatively small number of counts but with neighbouring enriched regions may not be detected by the mixture model but would be detected by the MRF model.
Similarly to before, Figure \ref{fig:ChromHMM_2} shows the results of {\tt ChromHMM} using a 4-state hidden Markov model on the enrichment profile given by MRF and the mixture model, each at a 5\% FDR. %For each method, the data from both proteins is jointly modelled by {\tt ChromHMM}.
The emission probabilities (left) show how, for all analyses, two of the three states explain most of the enrichment pattern in the identified lists. The relative fold enrichment plots (right) show how these two states seem to be mostly enriched with TSS and CpGIsland features for both methods. Together with the results in Table \ref{Table:realdata2}, one can conclude that by taking into account Markov properties while combining replicates, many more regions are found at the same FDR, and that these regions are of the same nature as those found by the mixture model.
TABLE \ref{Table:realdata2} and FIGURE \ref{fig:ChromHMM_2} ABOUT HERE.

It is expected that p300 and CBP should have roughly the same number of binding sites (\citealp{bao13}). In this case, one can impose this constraint in the model, as discussed in section \ref{section:same p}. Table
\ref{Table:realdata3} reports the number of enriched and
differentially bound regions for chromosome 21 under the assumption that the two proteins, CBP and
p300, have the same number of binding sites under the same condition (here the time point). In \citet{bao13}, we show how this approach helps to account for the different IP efficiencies of individual experiments, particularly in cases where there are no technical replicates and therefore it is more difficult to give an accurate estimate for the proportion of binding sites.
TABLE \ref{Table:realdata3} ABOUT HERE.

\vspace{-1cm}
\section{Conclusion}
\label{section:Discussion}
In this paper, we propose a one-dimensional Markov random field model for the analysis of ChIP-seq data. Our model can be viewed as a
hidden Markov model where the initial distribution is the stationary
distribution. As such, we follow the literature on existing HMM-based models, such as BayesPeak (\citealp{spyrou09}) and iSeq (\citealp{mo12}). Similarly to these models, we capture the spatial dependencies of local bins by an assumption of first-order Markov properties. Differently to these methods, we develop a joint model for multiple ChIP-seq experiments under general experimental designs, such as experiments with replicates and different antibodies. The resulting joint model is expected to lead to  a more robust detection of enriched and differentially bound regions. Furthermore, similarly to our previously developed mixture model (\citealp{bao13}), we show how a priori knowledge of the same number of binding sites for different proteins can also be added to the model, in order to better account for the different ChIP efficiencies of individual experiments.  Finally, we advocate the use of zero-inflated distributions for the background distribution, as these better account for the large number of zeros in the data.

In an extensive simulation study, we show how the proposed Markov random field model is in general superior to both iSeq and BayesPeak, as it achieves a lower false non-discovery rate at the same false discovery rate. When the data are generated from the same model used by iSeq, i.e. an Ising model with one parameter and a Poisson mixture, the methods perform similarly well, but RFM performs better than iSeq and BayesPeak under more general mixture distributions and model assumptions. Finally, we present an analysis on real data for the detection of histone modifications of two transcriptional activators from eight ChIP-sequencing experiments, including technical replicates with different IP efficiencies.
\vspace{-1cm}
\section{Software}
\label{section:Software}
The method is available in the R package {\tt enRich}, which can be currently downloaded from \url{http://people.brunel.ac.uk/~mastvvv/Rcode/enRich_2.0.tar.gz}.
\vspace{-1cm}
\section{Supplementary Material}
\label{section:Supplementary}
Supplementary material is available online at
\href{http://biostatistics.oxfordjournals.org}%
{http://biostatistics.oxfordjournals.org}.
\vspace{-1cm}
\section*{Acknowledgments}
  This work was supported by the Biotechnology and Biological Sciences Research Council [BB/H017275/1 to Y.B.]; the European Commission 7th Framework Program GEUVADIS [project nr. 261123 to P.'t H.]; and the Centre for Medical Systems Biology within the framework of the Netherlands Genomics Initiative/Netherlands Organisation for Scientific Research.\\
{\it Conflict of Interest}: None declared.
\vspace{-1cm}
\bibliographystyle{chicago}
\bibliography{mrf}
%\newpage
\begin{center}
\begin{figure}[H]
\includegraphics[width=7cm,height=8cm,angle=270]{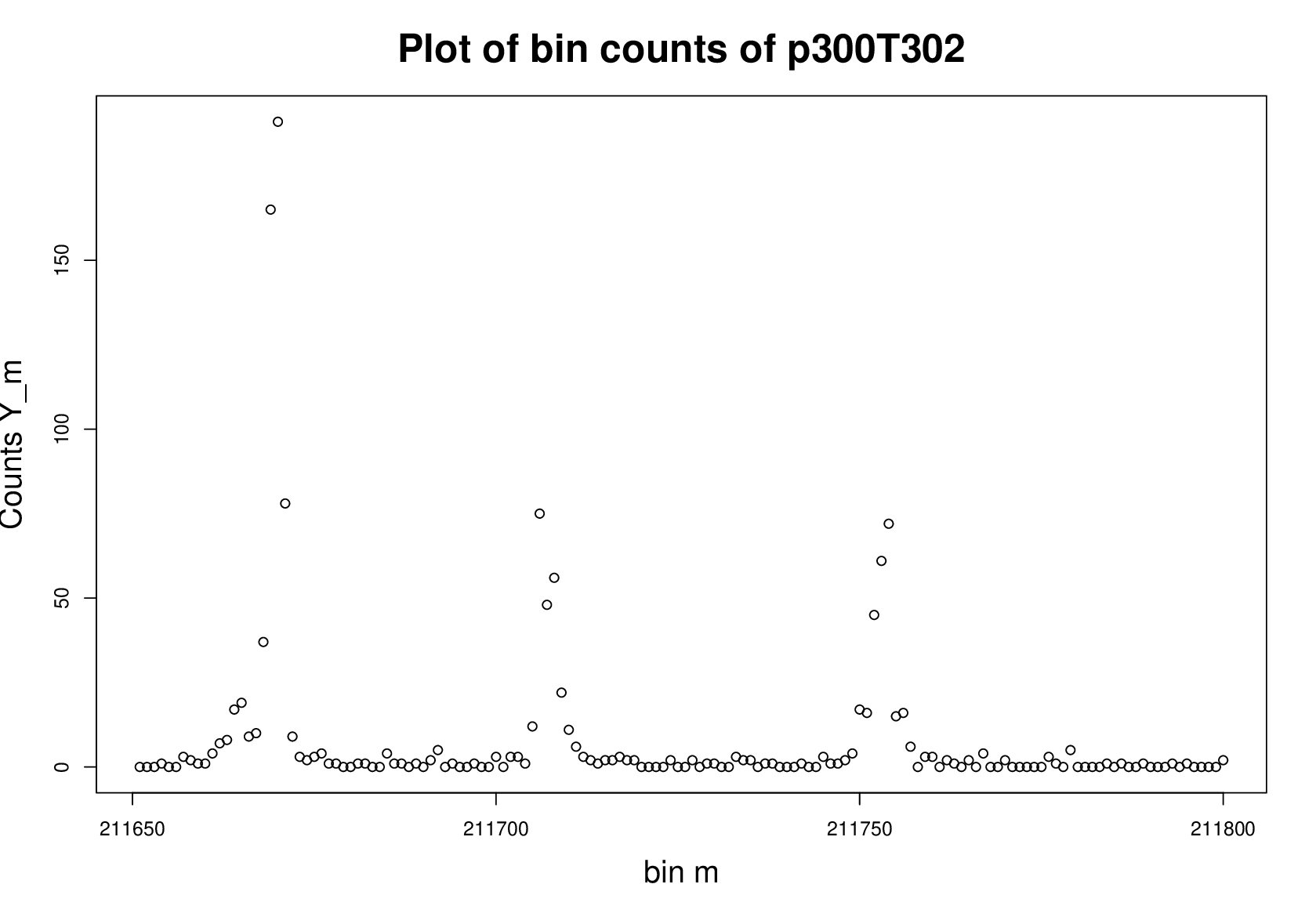}
\includegraphics[width=7cm,height=8cm,angle=270]{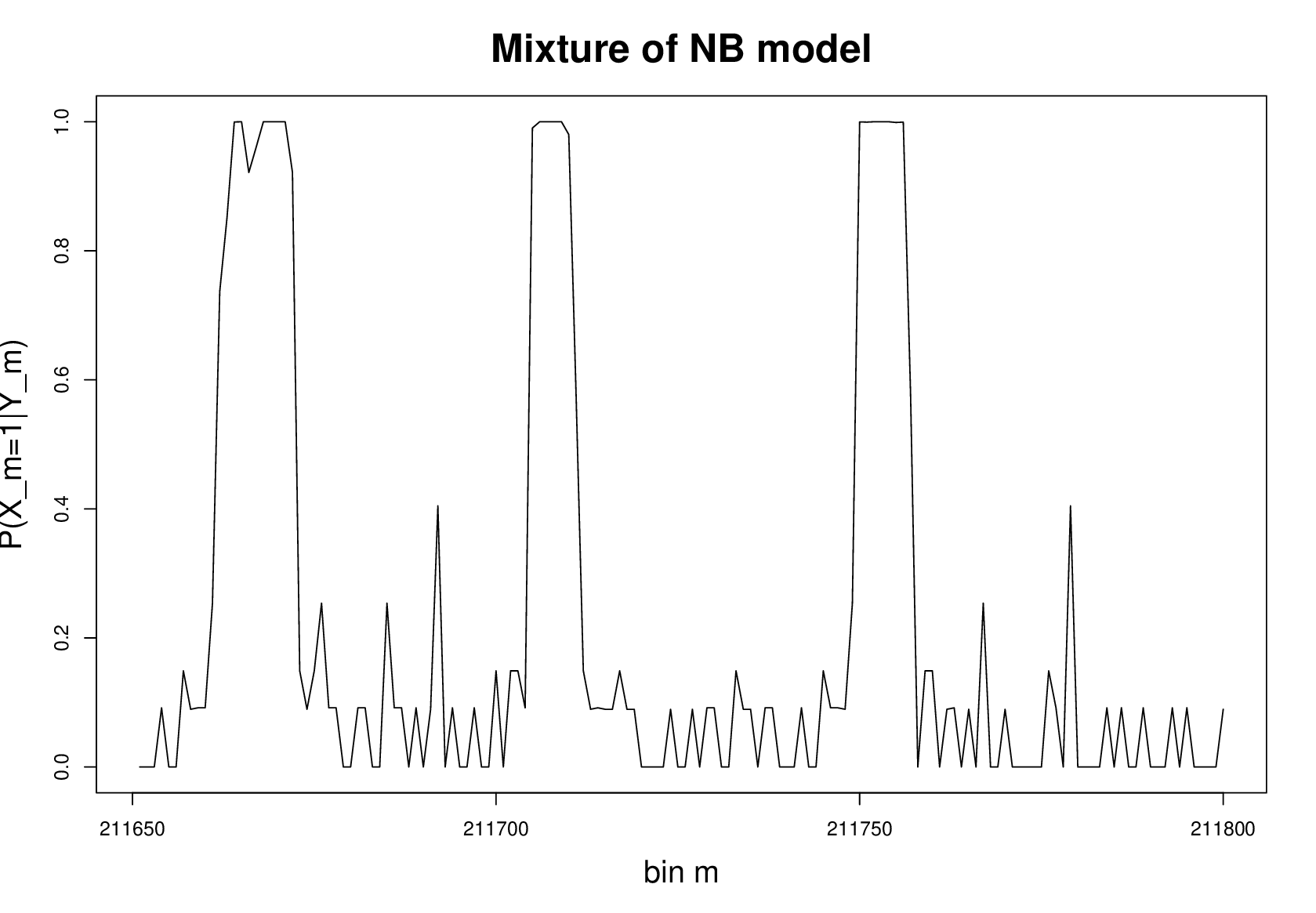}
\caption{Plots of bin counts $Y_m$ versus bin $m$ (left) for bins of size 200bp on a region of chromosome 1 for
the p300T302 experiment, and $P(X_m=1|Y_m, {\hat \theta})$ versus bin $m$ (right) under mixture of NB model.}\label{fig:markov_plot}
\end{figure}
\begin{figure}[H]
\includegraphics[width=7cm,height=9cm, angle=270]{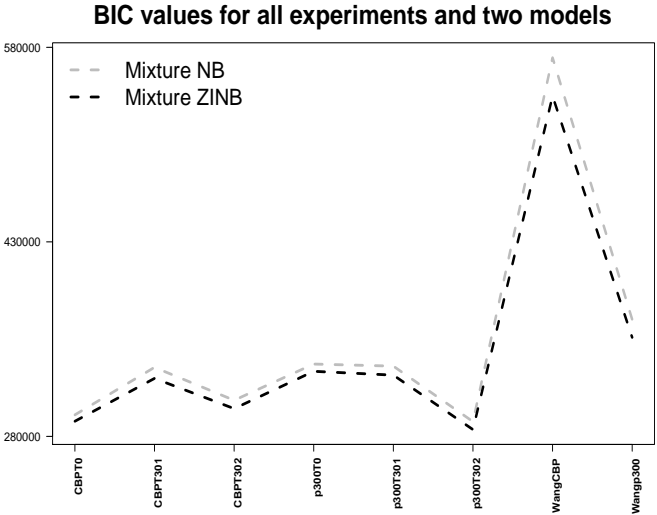}
\caption{BIC values for mixture of two NB distributions (dashed grey line) and mixture of
zero-inflated NB distribution for the background and NB distribution for the signal (dashed black line).
}\label{fig:BIC}
\end{figure}
\begin{figure}[H]
\includegraphics[width=7cm,height=9cm, angle=270]{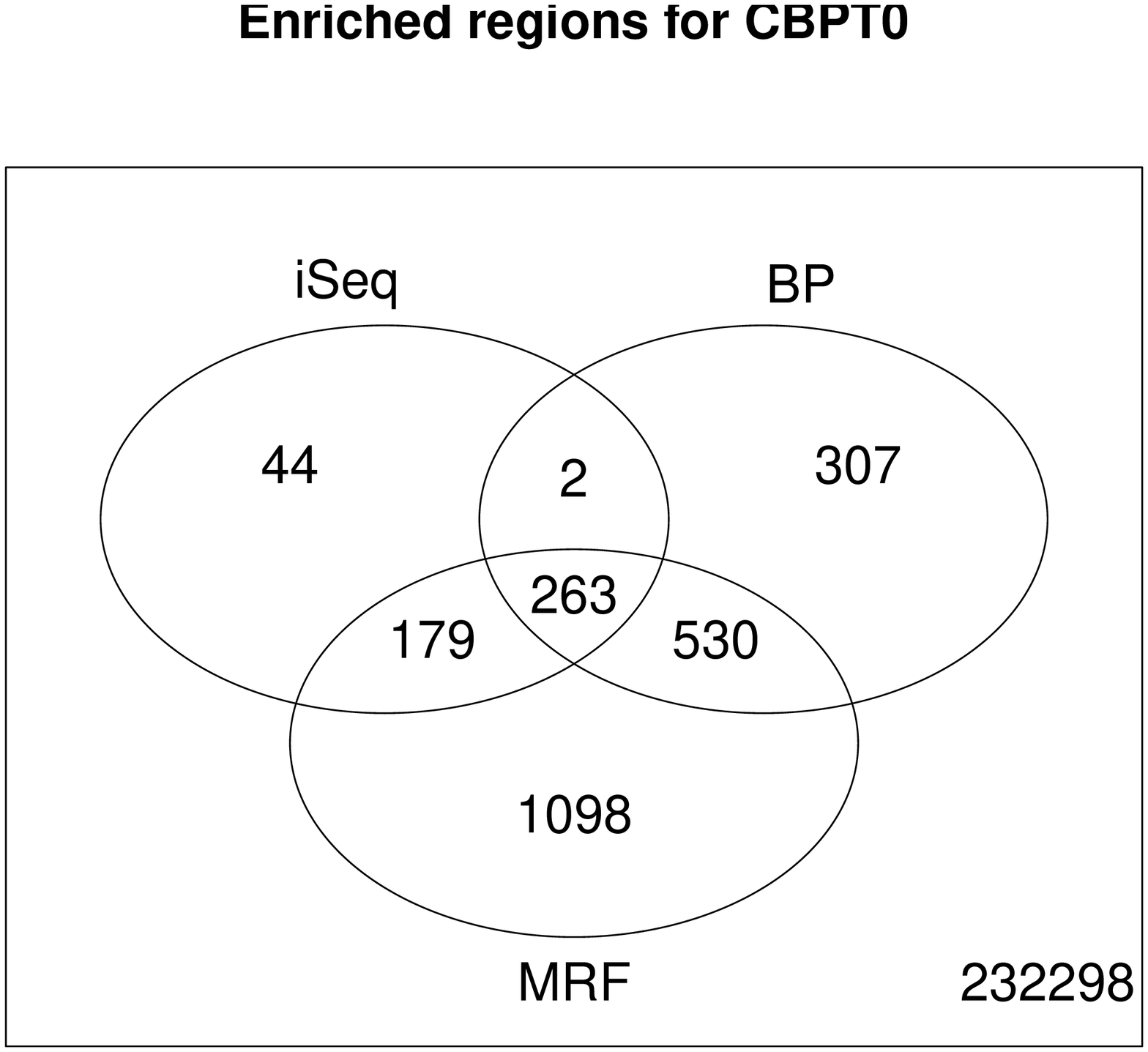}
\includegraphics[width=7cm,height=9cm, angle=270]{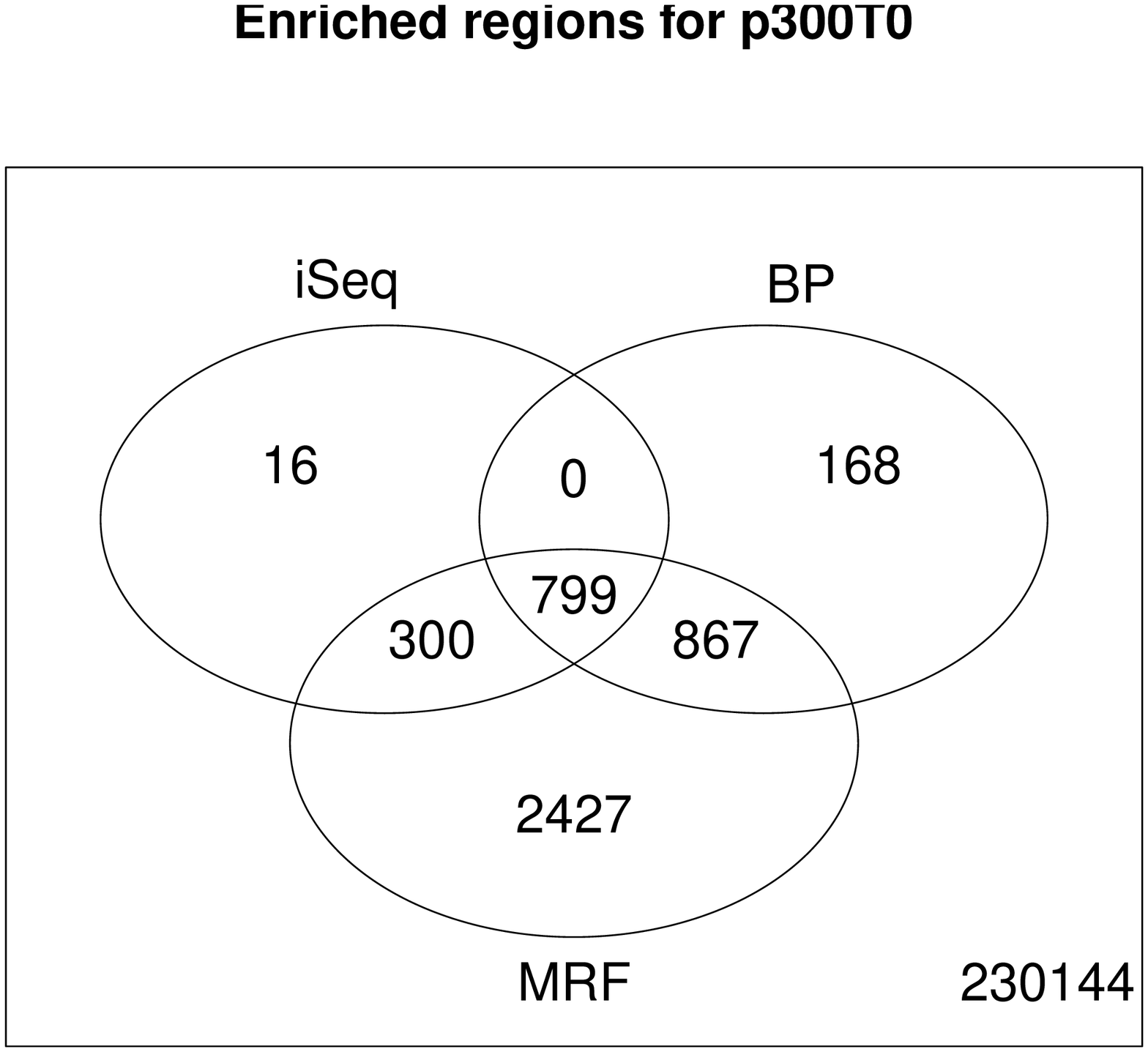}
\caption{Venn diagrams of the number of enriched regions detected by MRF, iSeq and BayesPeak (BP) at the 5\% FDR, for the CBPT0 and p300T0 experiments.} \label{fig:venndiagram}
\end{figure}
\begin{figure}[H]
\includegraphics[width=7cm,height=6cm]{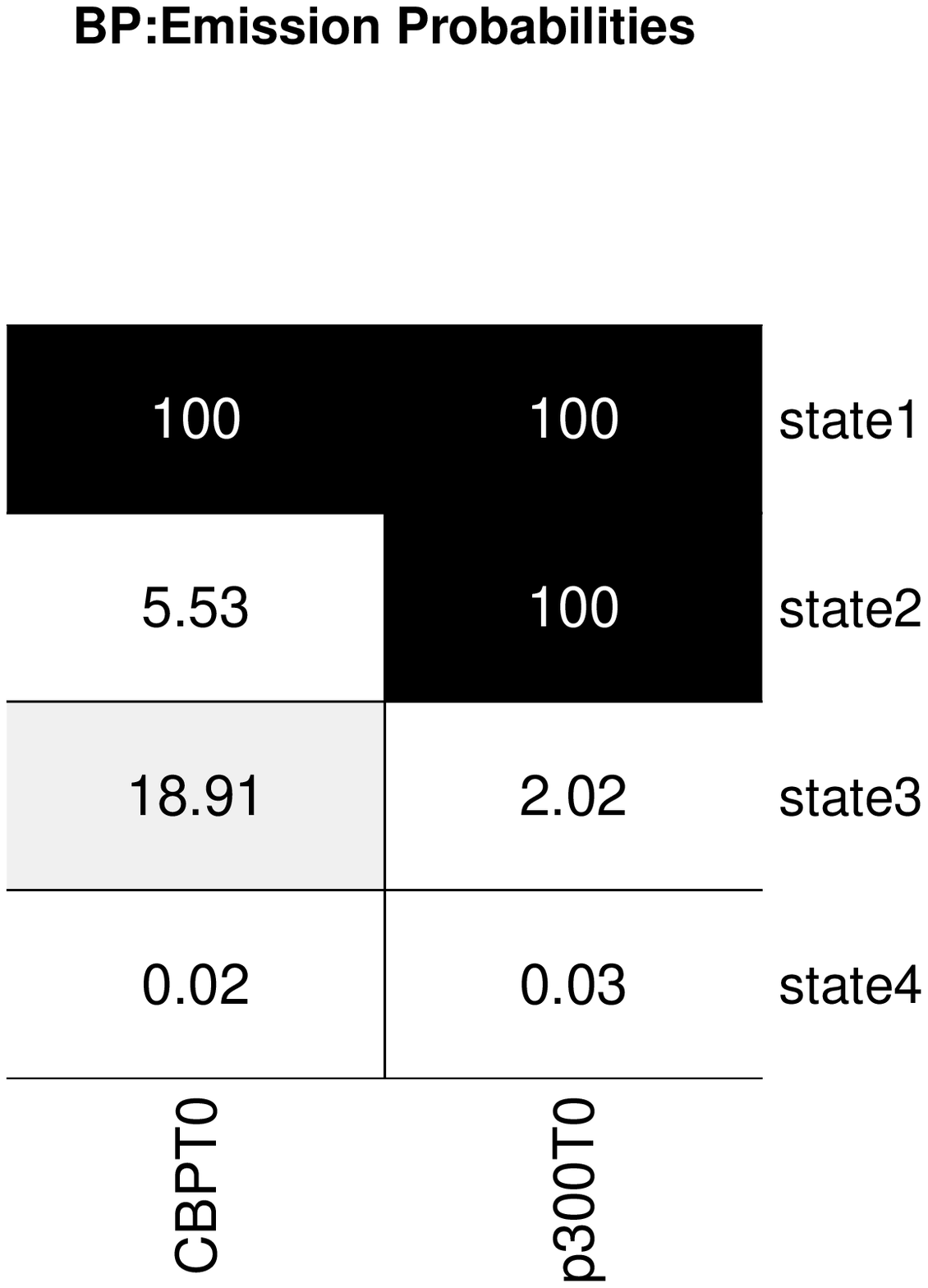}
\includegraphics[width=8cm,height=6cm]{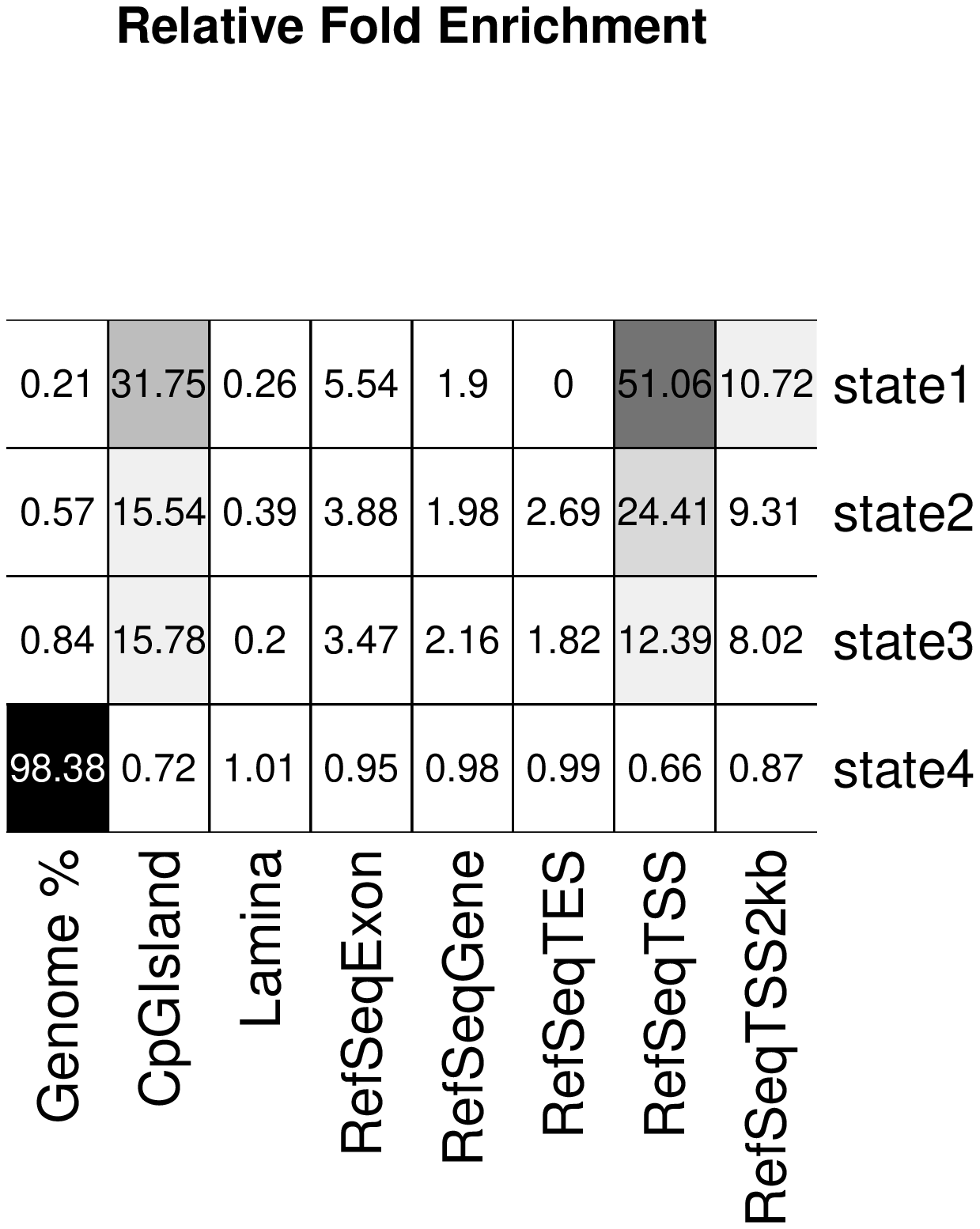}
\includegraphics[width=7cm,height=6cm]{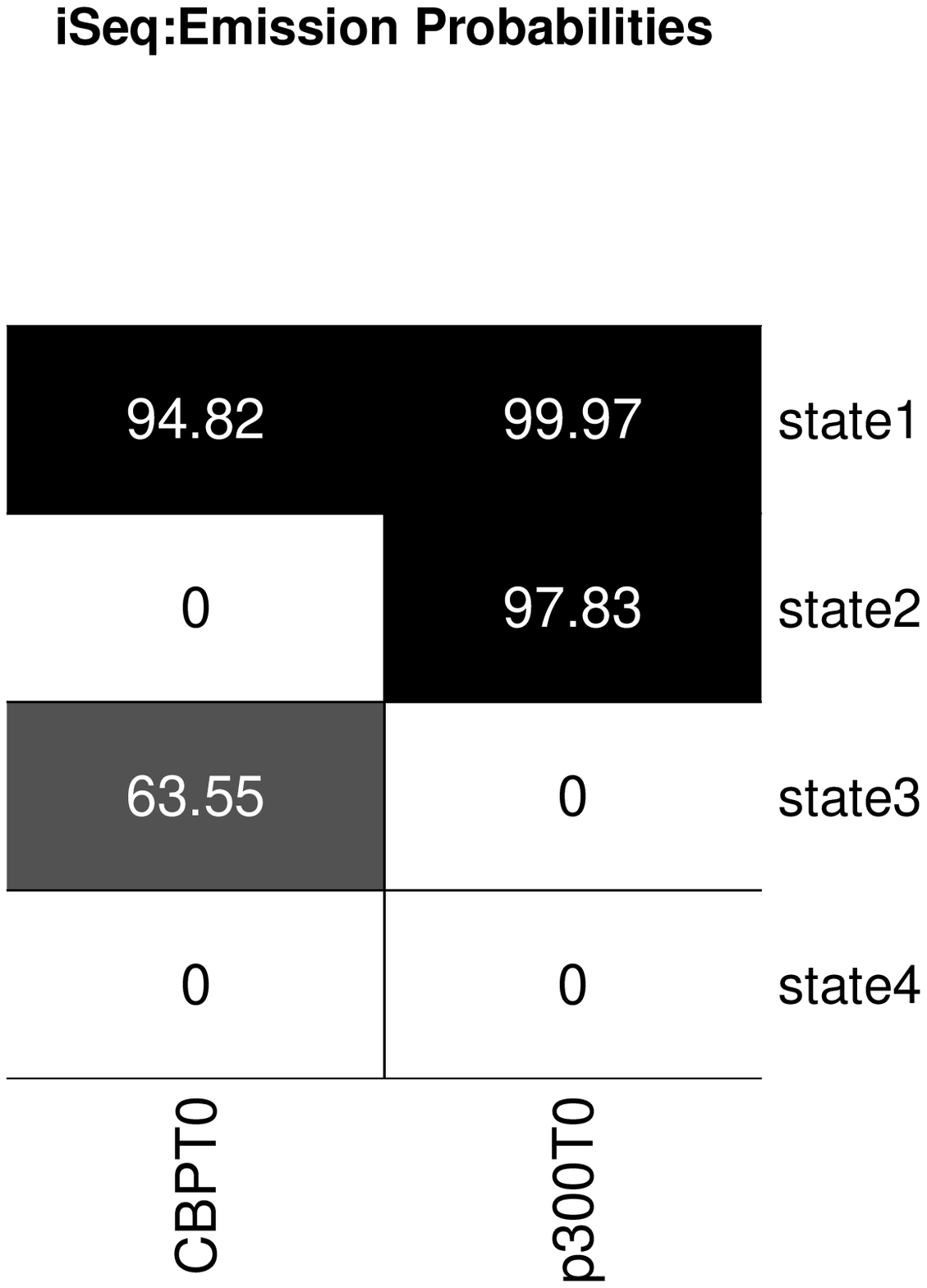}
\includegraphics[width=8cm,height=6cm]{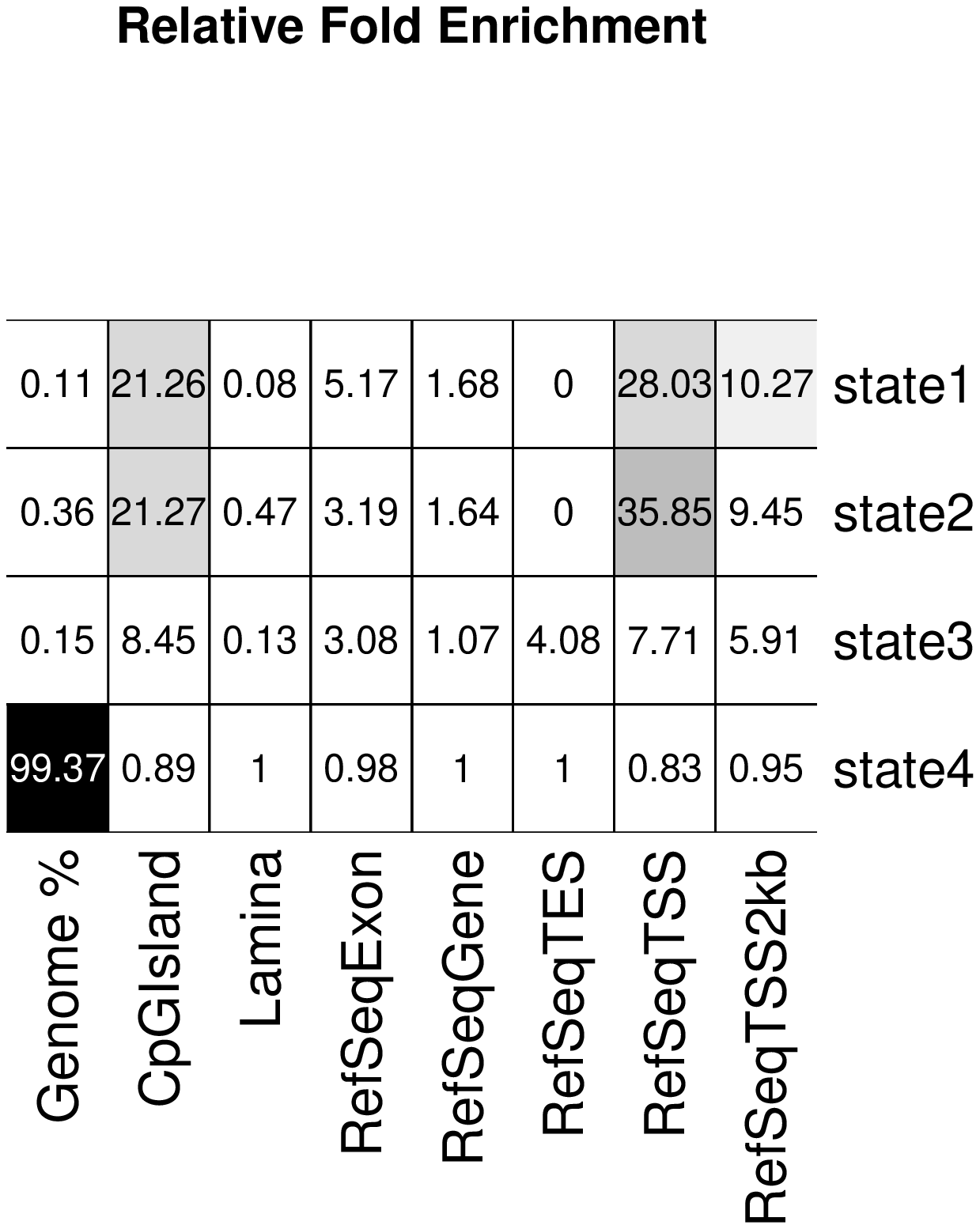}
\includegraphics[width=7cm,height=6cm]{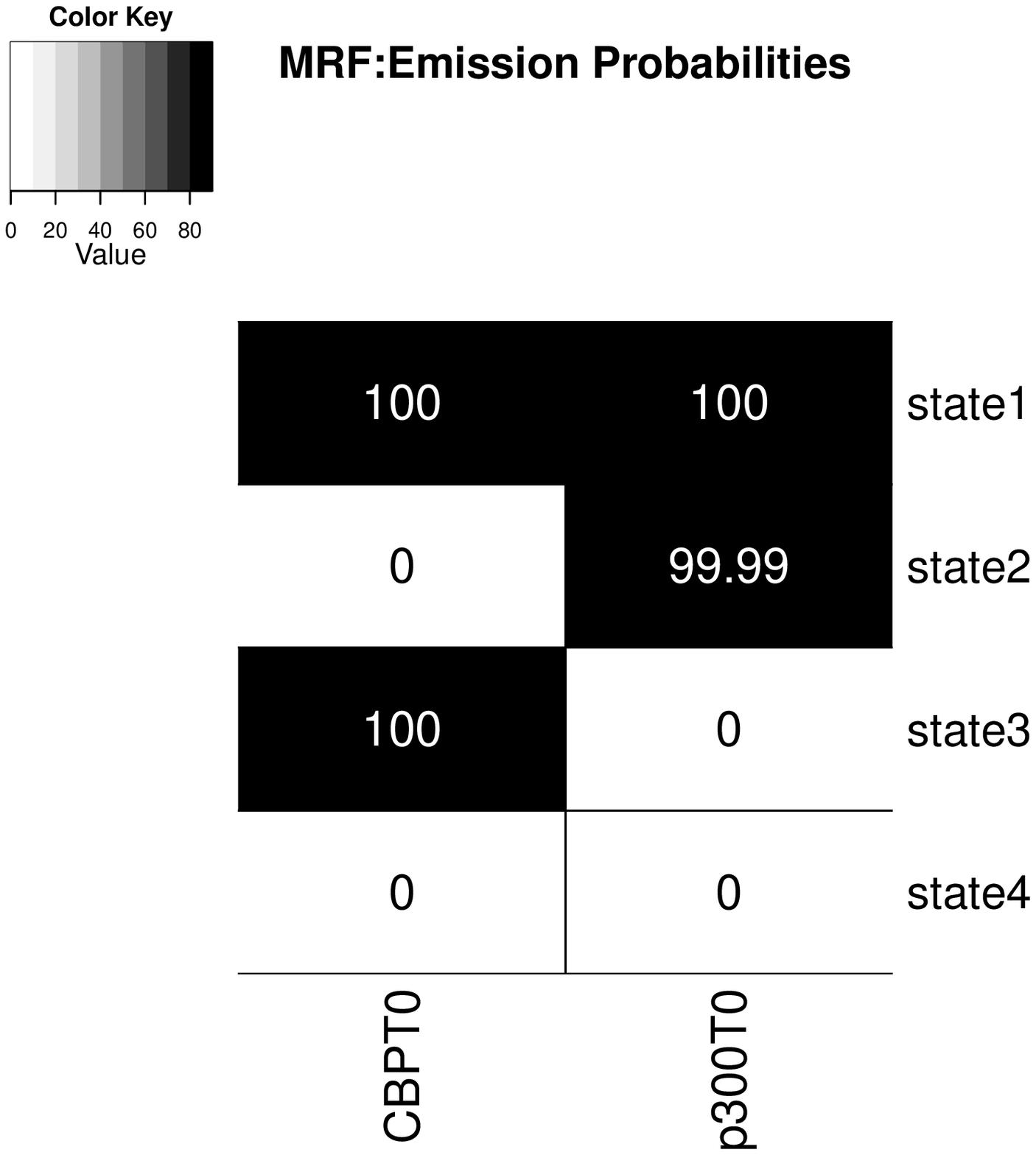}
\includegraphics[width=8cm,height=6cm]{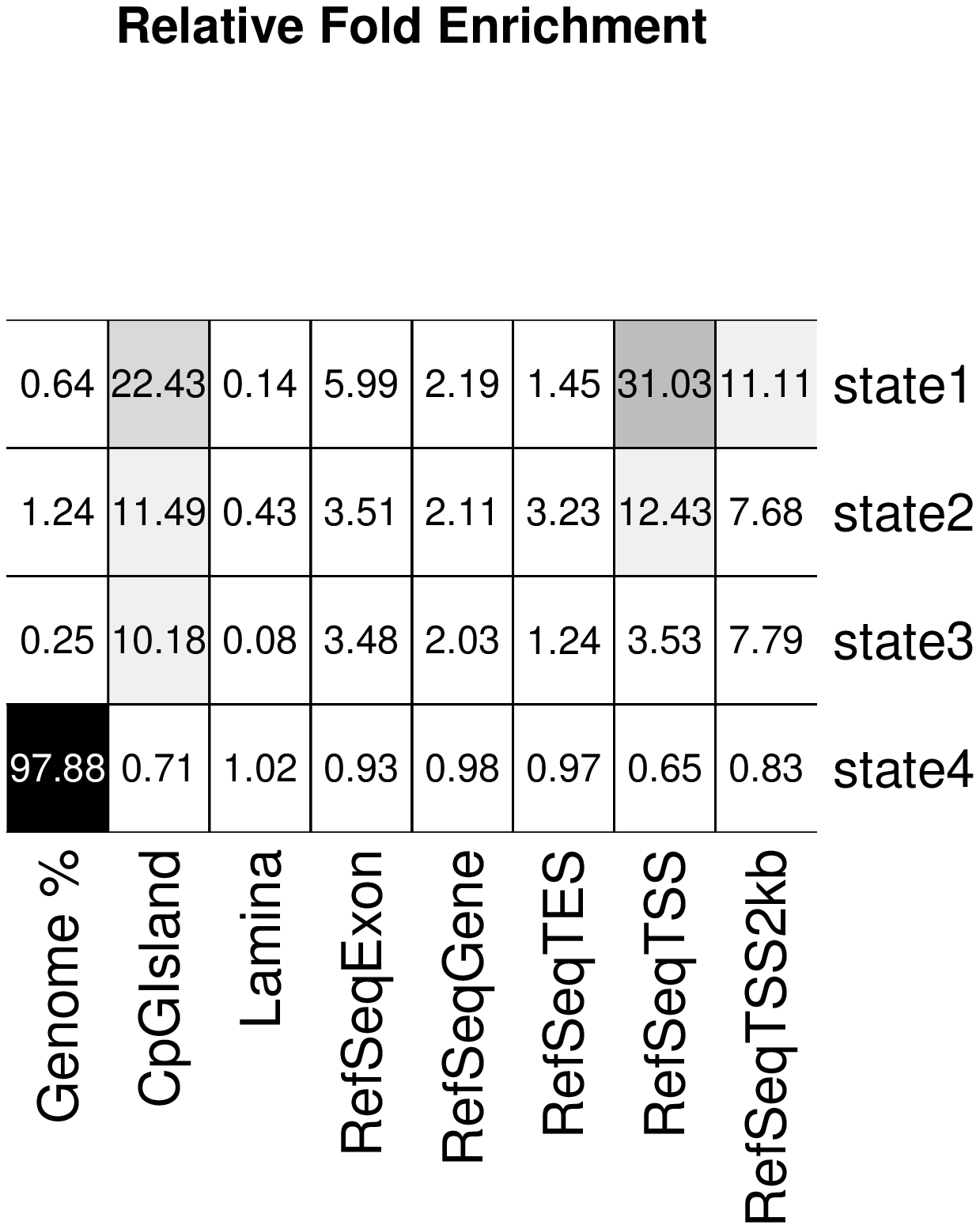}
\caption{Validation of the enriched bins detected by BayesPeak (BP, top), iSeq (middle) and MRF (bottom) for CBPT0 and p300T0,
using ChromHMM with a 4-state hidden Markov model. The left plots show heatmaps of the probabilities (in \%) that the detected bins are enriched given each identified chromatin-state. The right plots show  the relative percentage of the genome represented by each chromatin state (column 1) and the relative fold enrichment for several types of annotation (columns 2-8).}\label{fig:ChromHMM_1}
\end{figure}
\begin{figure}[H]
\includegraphics[width=6cm,height=5.5cm]{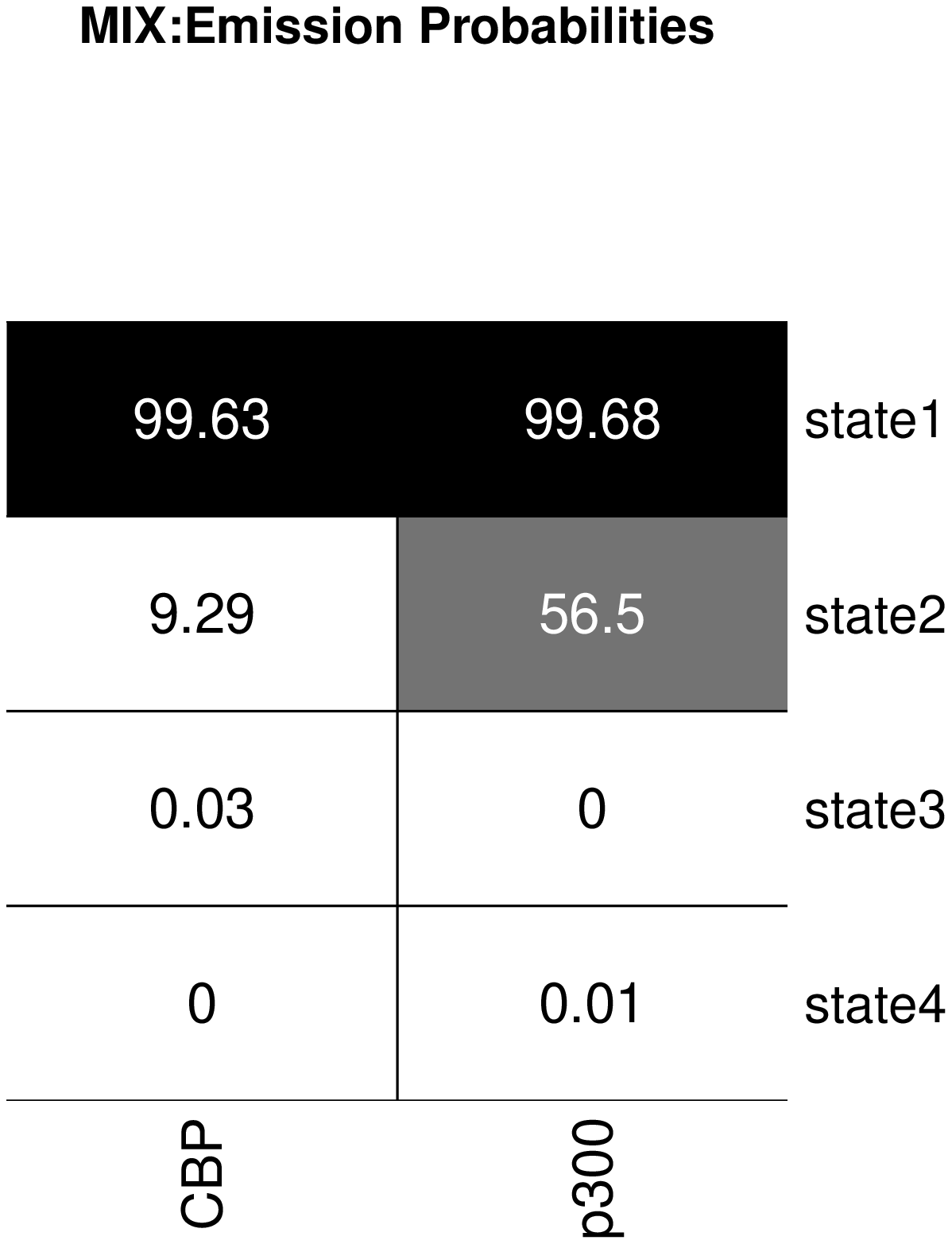}
\includegraphics[width=7cm,height=5.5cm]{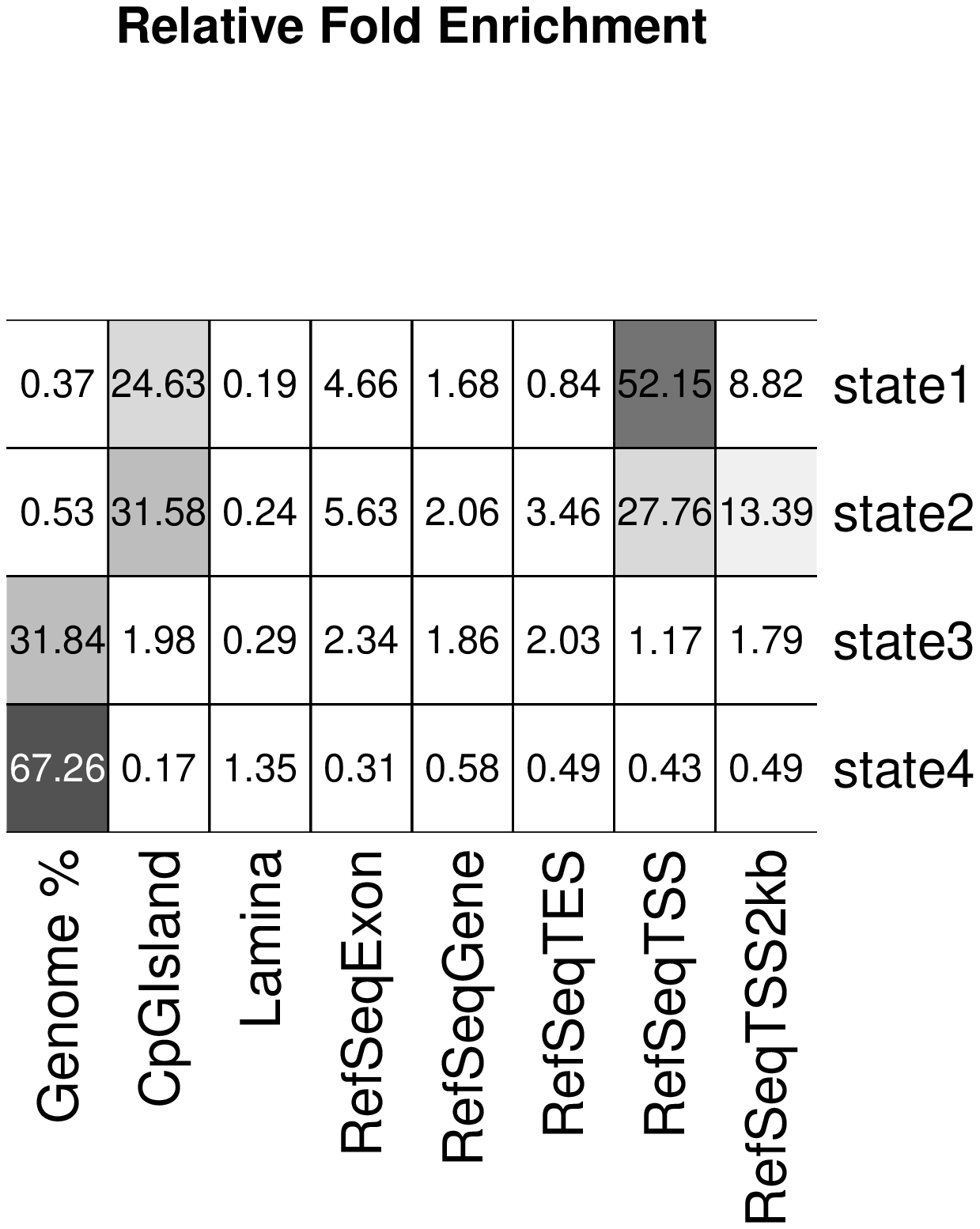}
\includegraphics[width=6cm,height=5.5cm]{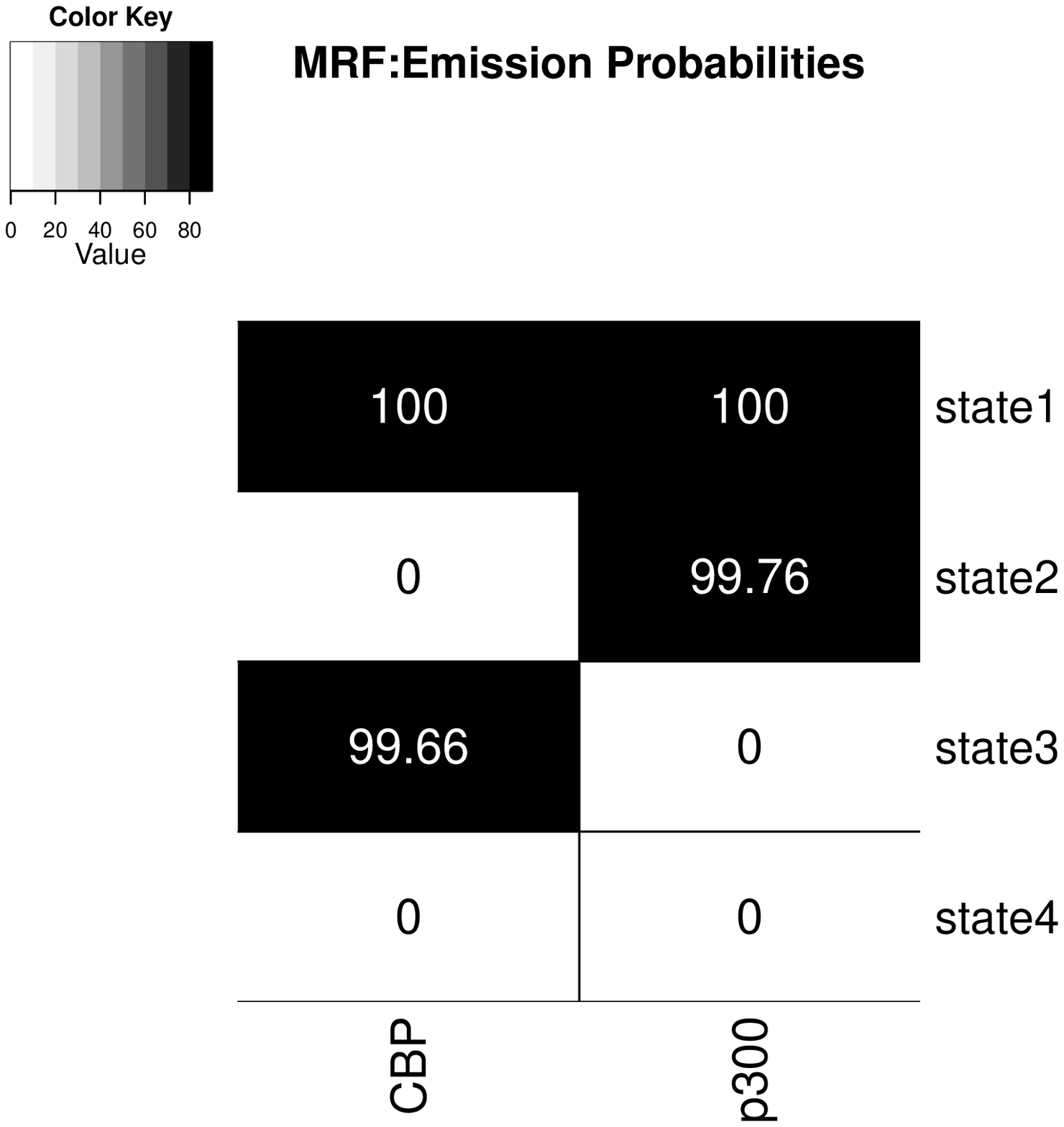}
\includegraphics[width=7cm,height=5.5cm]{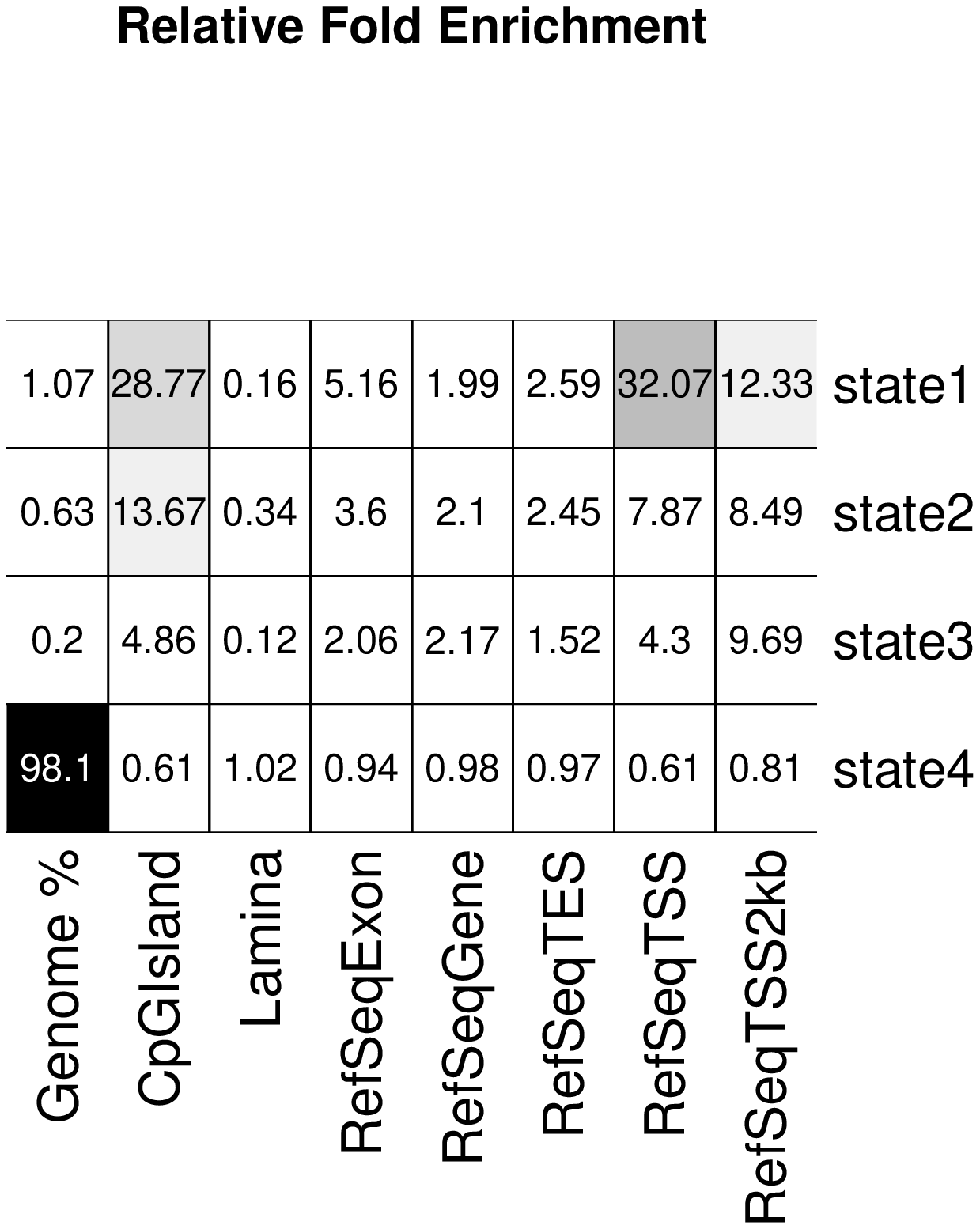}
\caption{Validation of the enriched bins detected by mixture model (MIX, top) and MRF (bottom) for technical replicates of CBP and p300 at time T30,  using a 4-state ChromHMM. The left plots show heatmaps of the probabilities (in \%) that the detected bins are enriched given each identified chromatin-state. The right plots show  the relative percentage of the genome represented by each chromatin state (column 1) and the relative fold enrichment for several types of annotation (columns 2-8).}\label{fig:ChromHMM_2}
\end{figure}
\begin{table}[H]
\caption{Conditional frequencies of enrichment given that the
previous bin is enriched or not, denoted by $f_{1|1}$ and
$f_{1|0}$ respectively. A region is called enriched or not using a latent mixture model at a 5\% FDR.}\label{Table:conditional_freq}
\resizebox{5cm}{!} {\begin{tabular}{lllllll}
\hline\\
Experiment&$f_{1|1}$&$f_{1|0}$\\
\hline\\
CBPT0 & 0.0909& 0.0002 \\
CBPT301 & 0.3004 & 0.0008\\
CBPT302 & 0.4924 & 0.0017\\
p300T0 & 0.3342 & 0.0021\\
p300T301 & 0.4015 & 0.0020\\
p300T302 & 0.5653 & 0.0027\\
Wang CBP & 0.4129 & 0.0014\\
Wang p300 & 0.2190 & 0.0005\\
\hline
\end{tabular}}
\end{table}
\end{center}
\begin{table}
\caption{Simulated count data is generated for $M=10000$ regions under four different scenarios. The table reports the average FNDR over 100 iterations, at a controlled FDR of 5\%, for MRF, iSeq and BayesPeak. The p-values show whether the MRF model has a significantly lower FNDR.}\label{Table:simulation1}
\resizebox{14cm}{!} {\begin{tabular}{|l|rr|rr|}\hline
&\multicolumn{2}{c|}{Less Efficient Experiment} & \multicolumn{2}{c|}{More Efficient Experiment}\\\hline
 &\multicolumn{4}{c|}{\bf Scenario 1: ZINB-NB mixture with $\tilde{q}_1+\tilde{q}_0\neq 1$ (as MRF).}\\
 \cline{2-5}
  &\multicolumn{2}{l|}{Signal: NB(1.38,2.07)}
  &\multicolumn{2}{l|}{Signal: NB(6.95,0.89)}\\
  &\multicolumn{2}{l|}{ Background: ZINB(0.66, 0.33, 2.01)}
  &\multicolumn{2}{l|}{ Background: ZINB(0.53, 0.36, 0.88)}\\
  &\multicolumn{2}{l|}{$(\tilde{q}_0,\tilde{q}_1)$=(0.002,0.940)}
  &\multicolumn{2}{l|}{$(\tilde{q}_0,\tilde{q}_1)$=(0.003,0.866) }\\ \hline
              &  FNDR   &   p-value  &   FNDR  & p-value    \\
MRF           & 0.0090 &   -        & 0.0020 &      -     \\
iSeq          & 0.0086 & 0.7778     & 0.0052 & $<2.2e-16$ \\
BayesPeak     & 0.0292 & $<2.2e-16$ & 0.0088 & $<2.2e-16$ \\
\hline
&\multicolumn{4}{c|}{\bf Scenario 2: Poisson-Poisson mixture with  $\tilde{q}_1+\tilde{q}_0 = 1$ (as iSeq).}\\
 \cline{2-5}
 &\multicolumn{2}{l|}{Signal: Poisson(1.5)}
 &\multicolumn{2}{l|}{Signal: Poisson(9.0)}\\
 &\multicolumn{2}{l|}{Background: Poisson(0.5)}
 &\multicolumn{2}{l|}{Background: Poisson(0.5)}\\
 &\multicolumn{2}{l|}{$\tilde{q}_1=1-\tilde{q}_0=0.98$}
 &\multicolumn{2}{l|}{$\tilde{q}_1=1-\tilde{q}_0=0.98$}\\ \hline
              &    FNDR &   p-value  &   FNDR     & p-value \\
MRF           & 0.0606 & -          & 3.79e-06  & -  \\
iSeq          & 0.0586 & 0.7661     &  1.04e-05 & 0.1073    \\
BayesPeak     & 0.4547 & $<2.2e-16$ & 0.2707    & $<2.2e-16$ \\ \hline
&\multicolumn{4}{c|}{\bf Scenario 3: Poisson-Poisson mixture with $\tilde{q}_1+\tilde{q}_0\neq 1$.}\\ \cline{2-5}
 &\multicolumn{2}{l|}{Signal: Poisson(3.0)}& \multicolumn{2}{c|}{Signal: Poisson(6.0)}\\
 &\multicolumn{2}{l|}{ Background: Poisson(0.5)}&\multicolumn{2}{c|}{Background: Poisson(0.2)}\\
 &\multicolumn{2}{l|}{$(\tilde{q}_0,\tilde{q}_1)=(0.02,0.5)$}
 &\multicolumn{2}{l|}{$(\tilde{q}_0,\tilde{q}_1)=(0.02,0.5)$}\\ \hline
              &    FNDR &   p-value  &   FNDR   & p-value \\
MRF           & 0.0225 & -          & 0.0011  & -  \\
iSeq          & 0.0287 & $<2.2e-16$ & 0.0016  & $1.737e-12$ \\
BayesPeak     & 0.0299 & $<2.2e-16$ & 0.0200  & $<2.2e-16$ \\ \hline
&\multicolumn{4}{c|}{\bf Scenario 4: ZINB-NB mixture with $\tilde{q}_1+\tilde{q}_0=1$.}\\ \cline{2-5}
&\multicolumn{2}{l|}{Signal: NB(3.0, 1.0)}
&\multicolumn{2}{l|}{Signal: NB(6.0, 1.0)}\\
&\multicolumn{2}{l|}{Background: ZINB(0.5, 0.5,0.5)}
&\multicolumn{2}{l|}{Background: ZINB(0.5, 0.5,0.5)}\\
&\multicolumn{2}{l|}{$(\tilde{q}_0,\tilde{q}_1)=(0.02, 0.98)$}
&\multicolumn{2}{l|}{$(\tilde{q}_0,\tilde{q}_1)=(0.02, 0.98)$}\\ \hline
              &    FNDR &   p-value  &   FNDR   & p-value \\
MRF           & 0.0168 & -          & 0.0039  & -  \\
iSeq          & 0.2903 & $<2.2e-16$ & 0.1874  & $<2.2e-16$ \\
BayesPeak     & 0.4100 & $<2.2e-16$ & 0.4310  & $<2.2e-16$ \\
\hline
\end{tabular}}
\end{table}
\begin{table}
\caption{Simulated count data are generated for $M=10000$ regions and for two replicates from a ZINB-NB under two different scenarios (with and without Markov property). The table reports the average FNDR over 100 iterations, at a controlled FDR of 5\%, for MRF and a mixture model. The p-values show whether the MRF model has a significantly lower FNDR.}
\label{Table:simulation2}
\resizebox{14cm}{!}{\begin{tabular}{|l|rr|rr|}\hline
&\multicolumn{2}{c|}{Less Efficient Experiment} & \multicolumn{2}{c|}{More Efficient Experiment}\\\hline
&\multicolumn{4}{c|}{\bf Scenario 5: multiple experiments and Markov property.}\\ \cline{2-5}
&\multicolumn{2}{l|}{Rep 1 -- Signal: NB(2.738, 1.548)}
&\multicolumn{2}{l|}{Rep 1 -- Signal: NB(3.797, 1.139)}\\
&\multicolumn{2}{l|}{Rep 2 -- Signal: NB(5.991, 0.957)}
&\multicolumn{2}{l|}{Rep 2 -- Signal: NB(7.392, 0.955)}\\
&\multicolumn{2}{l|}{Rep 1 -- BG: ZINB(0.634, 0.430, 2.322)}
&\multicolumn{2}{l|}{Rep 1 -- BG: ZINB(0.656, 0.393, 3.014)}\\
&\multicolumn{2}{l|}{Rep 2 -- BG: ZINB(0.481, 0.477, 1.246)}
&\multicolumn{2}{l|}{Rep 2 -- BG: ZINB(0.486, 0.395, 1.061) }\\
&\multicolumn{2}{l|}{$(\tilde{q}_0,\tilde{q}_1)=(0.003,0.839)$} &\multicolumn{2}{c|}{$(\tilde{q}_0,\tilde{q}_1)=(0.003,0.830)$}\\
\hline
&    FNDR &   p-value &   FNDR   & p-value  \\
MRF & 0.0011 & -     & 0.0008  & -     \\
Mixture      & 0.0072 & $<2.2e-16$ & 0.0057  & $<2.2e-16$\\
\hline
&\multicolumn{4}{c|}{\bf Scenario 6: multiple experiments and no Markov property.}\\ \cline{2-5}
&\multicolumn{2}{l|}{Rep 1 -- Signal: NB(2.738, 1.548)}
&\multicolumn{2}{l|}{Rep 1 -- Signal: NB(3.797, 1.139)}\\
&\multicolumn{2}{l|}{Rep 2 -- Signal: NB(5.991, 0.957)}
&\multicolumn{2}{l|}{Rep 2 -- Signal: NB(7.392, 0.955)}\\
&\multicolumn{2}{l|}{Rep 1 -- BG: ZINB(0.634, 0.430, 2.322)}
&\multicolumn{2}{l|}{Rep 1 -- BG: ZINB(0.656, 0.393, 3.014)}\\
&\multicolumn{2}{l|}{Rep 2 -- BG: ZINB(0.481, 0.477, 1.246)}
&\multicolumn{2}{l|}{Rep 2 -- BG: ZINB(0.486, 0.395, 1.061)}\\
&\multicolumn{2}{l|}{$p(X=1)=0.017$}
&\multicolumn{2}{l|}{$p(X=1)=0.020$}\\
\hline
                   &    FNDR &   p-value &   FNDR   & p-value \\
MRF & 0.0073 & -    & 0.0058  & -        \\
Mixture      & 0.0073 & $0.5001$ & 0.0058  & $0.6738$\\
\hline
\end{tabular}}
\end{table}
\begin{table}
\caption{Number of enriched regions identified by MRF, iSeq and BayesPeak for four single experiments.}\label{Table:realdata1}
\resizebox{10cm}{!}
{\begin{tabular}{|l|rrrr|}\hline
Method   & CBPT0 & p300T0& WangCBP & Wangp300\\%  Bound only by CBP & Bound only by p300\\
\cline{1-5}
MRF      & 2073 & 4393 & 1443 &639\\%70 & 557 \\%& 2145 &465 & 100\% & 4487 &1079 & 100\% \\
iSeq     &  488 & 1115 & 1126 & 326\\%-  & -   \\%&  501 &287 & 62\% & 1129 &805 & 75\%\\%
BayesPeak& 1102 & 1834 & 603 &576\\ \hline%-  & -   \\%& 1995 &465 &100\% & 2812 &1079 & 100\% \\ %
overlap  & 263  &  799 &  396 & 190\\%-  & -\\%& 290 &465 & - & 963 &1079 &-\\%
\hline
\end{tabular}}
\end{table}
\begin{table}
\caption{Number of enriched and differentially bound regions identified by MRF and ZINB-NB mixture
model from technical replicates of CBP and p300 at time 30.
}\label{Table:realdata2}
\resizebox{10cm}{!} {\begin{tabular}{|l|rr|rr|}\hline
       & \multicolumn{2}{c|}{Enriched regions}&\multicolumn{2}{c|}{Differentially bound regions}\\
Method & CBPT30 & p300T30 & only CBP & only p300\\
\hline
MRF           & 2977  & 3970   &  69   &  347 \\
Mixture &  981  & 1848   &  29   &  395 \\ \hline
overlap       &  971  & 1823   &  4   &   78  \\
\hline
\end{tabular}}
\end{table}
\begin{table}
\caption{Number of enriched and differentially regions identified by MRF under the assumption that the two proteins, CBP and
p300, have the same number of binding sites at the same time point.}\label{Table:realdata3}
\resizebox{12cm}{!}
{\begin{tabular}{|l|rr|rr|} \hline
& \multicolumn{2}{c|}{Enriched regions}&\multicolumn{2}{c|}{Differentially bound regions}\\
                  & CBP     & p300         & only CBP & only p300\\ \hline
CBPT0 vs p300T0   & 1606&   3842   &              23   &  336 \\
CBPT30 vs p300T30 & 3146  &   4123   &              62   &  365 \\
WangCBP vs Wangp300 &  1426 &   643   &             376   &  23  \\ \hline
\end{tabular}}
\end{table}
\end{document}